\documentclass[12pt]{article}
\pdfoutput=1
\usepackage{graphicx,psfrag,epsf,color}
\usepackage{amsmath,amssymb,amsfonts}
\usepackage{array}
\usepackage{cite}
\usepackage{rotating}
\usepackage{slashed, cancel}
\usepackage{xcolor}
\usepackage[caption=false]{subfig}
\usepackage{graphicx}
\usepackage{mathrsfs}  
\usepackage{tabularx}
\usepackage{multirow} \usepackage{array}
\usepackage{floatrow}
\usepackage{physics}

\usepackage{color,xcolor}
\usepackage{lscape}
\usepackage[colorlinks,citecolor=blue,urlcolor=blue,linkcolor=blue]{hyperref}

\newcolumntype{C}{>{\centering\arraybackslash}X}
\numberwithin{equation}{section}
\begin{document}
\allowdisplaybreaks

\begin{titlepage}

\begin{flushright}
{\small
CERN-TH-2024-206\\
SI-HEP-2024-26\\
P3H-24-093\\
December 18, 2024
}
\end{flushright}

\vskip1cm
\begin{center}
{\Large \bf\boldmath $B_c \to \eta_c$ form factors at large recoil: \\[0.4em] 
Interplay of soft-quark and soft-gluon dynamics
}
\end{center}

\vspace{0.5cm}
\begin{center}
Guido~Bell$^a$, Philipp B\"oer$^b$, Thorsten Feldmann$^a$, Dennis Horstmann$^a$ \\ and Vladyslav Shtabovenko$^a$ \\[6mm]

{\it $^a$Theoretische Physik 1, Center for Particle Physics Siegen, \\
Universit\"at Siegen, 57068 Siegen, Germany}\\[0.3cm]

{\it $^b$CERN, Theoretical Physics Department, CH-1211 Geneva 23, Switzerland}
\end{center}

\vspace{0.6cm}
\begin{abstract}
\vskip0.2cm\noindent

We perform an all-order analysis of double-logarithmic corrections to the so-called soft-overlap contribution to heavy-to-light transition form factors at large hadronic recoil. Specifically, we study $B_c \to \eta_c$ transitions within a perturbative non-relativistic framework, treating both the bottom and charm quarks as heavy with the hierarchy $m_b \gg m_c \gg\Lambda_{\rm QCD}$. Our diagrammatic analysis shows that double-logarithmic corrections arise from two distinct sources: Exponentiated soft-gluon effects described by standard Sudakov factors, and rapidity-ordered soft-quark configurations, leading to implicit integral equations, which so far have only been studied in the context of energetic muon-electron backward scattering. We find that the all-order structure of the double logarithms is governed by a novel type of coupled integral equations, which encode the non-trivial interplay between these two effects. Whereas a closed-form solution to these equations is currently unknown, we present useful iteration formulas, and derive the asymptotic behaviour of the soft-overlap form factor for infinitely large recoil energies, showing that the Sudakov suppression is somewhat weakened by the intertwined soft-quark and soft-gluon corrections. In a broader context, our findings shed light onto the physical origin and mathematical structure of endpoint divergences arising from soft-collinear factorization and the related Feynman mechanism for power-suppressed hard exclusive processes.

\end{abstract}

\end{titlepage}

\section{Introduction}
\label{sec:Intro}

In the absence of evidence for physics beyond the Standard Model from direct searches at the high-energy frontier, weak decays of $B$-mesons remain among the most promising processes to test the flavour sector of the Standard Model (SM) and to find indirect hints of ``new physics''. To reveal the short-distance dynamics underlying flavour transitions in the SM or beyond, however, one has to deal with the complicated QCD dynamics involved in hadronic processes. It is well known that exclusive decays of heavy $B$-mesons allow for a systematic expansion in inverse powers of the $b$-quark mass, which simplifies their analysis by reducing the number of independent hadronic input parameters in the heavy-mass limit.

For decays into light and energetic particles, the situation is complicated by the fact that the recoil energy $E$ is of the same order as the heavy-quark mass.  Specifically, appropriate QCD factorization theorems for charmless $B$-decays \cite{Beneke:1999br,Beneke:2001ev} are formulated in terms of \emph{convolutions} of short-distance kernels and hadronic matrix elements of non-local operators. A field-theoretical framework for the factorization of soft and energetic particles is provided by Soft-Collinear Effective Theory (SCET) \cite{Bauer:2000yr,Bauer:2001yt,Bauer:2002nz,Beneke:2002ph,Beneke:2002ni}. In SCET,  soft and energetic (or ``collinear'') modes are represented by separate field operators with a well-defined power counting in the effective Lagrangian. Moreover, soft-collinear interaction terms are multipole-expanded with respect to the typical wavelengths associated with each mode. The renormalization-group evolution of the effective soft and collinear operators can then be used to resum large logarithms appearing in the perturbative expansion of short-distance coefficient functions.

As in every effective field theory (EFT), the SCET framework allows one to systematically study power corrections, i.e.\ subleading terms in the $1/m_b$ (respectively $1/E$) expansion. In comparison to other EFTs like Heavy-Quark Effective Theory, for which power corrections are well studied, several subtleties make soft-collinear factorization much more involved at subleading power, see e.g.~\cite{Beneke:2019kgv,Bodwin:2023asf,Boer:2023yde}. Arguably, the most intricate complication is the fact that the convolution integrals in the factorization formulas are often not well-defined at the endpoints of the integration region. This seems to be a generic feature of the SCET formalism and is related to the underlying separation of Feynman integrals in momentum regions~\cite{Beneke:1997zp} and the use of dimensional regularization. As exclusive and energetic hadronic decays represent power-suppressed processes in SCET, these endpoint divergences present a major obstacle for improving the theoretical predictions for a wide class of phenomenologically relevant $B$-physics observables.

In recent years, SCET factorization theorems and the resummation of large logarithmic corrections at subleading power have been studied for a number of processes. Among these are threshold logarithms in Drell-Yan and Higgs production~\cite{Beneke:2018gvs,Beneke:2019mua,Beneke:2020ibj,Broggio:2023pbu}, $e^+e^-$ event-shape distributions~\cite{Moult:2018jjd,Moult:2019uhz,Beneke:2022obx}, the bottom-quark induced $h \to\gamma\gamma$ and $h \to gg$ decays~\cite{Liu:2019oav,Liu:2020tzd,Liu:2020wbn,Wang:2019mym,Liu:2022ajh}, exclusive and inclusive $B$-meson decays~\cite{Cornella:2022ubo,Feldmann:2022ixt,Hurth:2023paz} as well as virtual Compton scattering~\cite{Schoenleber:2024ihr}. Starting from a factorization theorem in terms of ``bare'' quantities, soft-collinear factorization could be re-established in some of these cases after an \emph{additive rearrangement} of endpoint configurations, based on so-called ``refactorization identities''~\cite{Boer:2018mgl,Liu:2019oav}. However, it has also been figured out that in SCET applications to hard-exclusive reactions more complicated situations may appear, where the refactorization conditions enter in a \emph{recursive way}~\cite{Boer:2018mgl,Bell:2022ott,Boer:2023tcs}. As a consequence, the endpoint-divergent contributions cannot be rearranged in an additive manner, and an all-order factorization description is currently unknown for these processes. We expect this feature to appear generically for hard-exclusive reactions at subleading power.

As some of us argued in \cite{Bell:2022ott}, the simple QED process of muon-electron scattering at large energies and exact backward kinematics provides a template case to study this recursive pattern of endpoint singularities. The associated large double logarithms can be traced back to rapidity-ordered and on-shell \emph{soft-lepton} configurations in specific ladder-type Feynman diagrams. Their all-order structure is governed by an implicit integral equation, which can be solved in terms of modified Bessel functions \cite{Gorshkov:1966qd}. As we will show in this work, the situation for exclusive $B$-meson decays is more complicated, and to shed light onto the physical origin and mathematical structure of endpoint divergences -- and the associated pattern of large logarithmic corrections -- we investigate the double-logarithmic series of heavy-to-light $B_c \to \eta_c$ form factors at large hadronic recoil. Specifically, we treat the bottom and charm quark as heavy quarks with a mass hierarchy $m_b \gg m_c \gg \Lambda_{\rm QCD}$, such that the mesons can be approximated as non-relativistic bound states. In this approximation, the charm-quark mass serves as an infrared regulator and the hadronic transition form factors become calculable within QCD perturbation theory~\cite{Bell:2005gw,Bell:2006tz,Bell:2008er}. Hence, this setup provides a well-defined framework for studying soft-collinear factorization for a power-suppressed exclusive $B$-decay amplitude. We also stress that this is not intended to provide a phenomenologically accurate description of the $B_c \to \eta_c$ transition (for a review see e.g.~\cite{QuarkoniumWorkingGroup:2004kpm}); it rather allows us to address conceptual aspects related to factorization in a well-defined perturbative setup.

Using QCD-based diagrammatic resummation techniques, we derive a system of coupled integral equations, whose solution resums the large double logarithms for the so-called soft-overlap form factor in the $B_c \to \eta_c$ transition to all orders in perturbation theory. In the Abelian limit, these equations have already been presented in~\cite{Boer:2023tcs}. Notably, we identify two intertwined origins of double-logarithmic corrections: First, similar to muon-electron backward scattering, rapidity-ordered and on-shell \emph{soft-quark} configurations give rise to a pattern of nested  double-logarithmic corrections. Second, on-shell \emph{soft-gluon} configurations exponentiate to standard Sudakov factors. The non-trivial interplay between these two mechanisms then leads to a novel type of coupled integral equations that we examine in detail in this work. Whereas we could not find a closed-form solution to these equations, we elaborate on some of its properties and provide compact iteration formulas that can be used to evaluate the double-logarithmic series to practically any truncation order. Moreover, its perturbative expansion allows for independent cross-checks with fixed-order computations, which we performed up to next-to-next-to-next-to-leading order (N$^3$LO).  Details on this three-loop calculation will be presented in a forthcoming publication. In addition, for the first time we derive the asymptotic behaviour of the soft-overlap form factor in the large-energy limit, which reveals that the overall Sudakov suppression associated with the flavour-changing weak-transition vertex is somewhat weakened by the resummation of the intertwined soft-quark and soft-gluon corrections. To our knowledge, this analysis provides the first all-order discussion of large-logarithmic corrections to the soft-overlap form factor and the related Feynman mechanism, whose description in the standard QCD factorization approach would be plagued by endpoint singularities. Our results therefore provide important constraints for an all-order factorization theorem that is yet to be developed.

The remainder of this article is organized as follows. In Sec.~\ref{sec:setup} we introduce the notation and conventions used throughout this paper. Sec.~\ref{sec:fixed-order} is dedicated to a detailed analysis of double-logarithmic corrections in fixed-order perturbation theory, where we identify the two underlying dynamical mechanisms and study their non-trivial interplay. We then extra\-polate these findings to higher orders and present the coupled integral equations that govern the all-order structure in Sec.~\ref{sec:all-order}. We analyze this novel class of relations in greater detail in Sec.~\ref{sec:integraleqs}, where we also derive the asymptotic behaviour of the soft-overlap form factor in the large-energy limit. A brief discussion in Sec.~\ref{sec:discussion} puts our results into a broader context, and draws a connection between the integral equations and the renormalization-group evolution kernel of distribution amplitudes for light mesons. We finally conclude in Sec.~\ref{sec:conclusion}.

\section{Setup and notation}
\label{sec:setup}

The QCD dynamics of non-relativistic (NR) bound states is well understood (see e.g.~\cite{Beneke:2013jia,Beneke:2024sfa}).  Here we work at leading power in the NR expansion, which implies that we deal with Coulombic quark-antiquark systems in the \emph{static approximation}. More specifically, the $B_c^-$ meson consists of a heavy $b$-quark with momentum $p_b^\mu = m_b v^\mu$ and a ``light'' (but non-relativistic) $c$-antiquark with $\ell^\mu = m_c v^\mu$. The spinor degrees-of-freedom of the heavy pseudo\-scalar $B_c^-$ meson are encoded in the Dirac projector $\mathcal{P}_{B} \propto (1+ \slashed{v})\gamma_5$, where $v^\mu$ is the four-velocity of the meson ($v^2 = 1$). Likewise, the light pseudo\-scalar $\eta_c$ meson consists of a $c$-quark with momentum $p_1^\mu = m_c v'^\mu$ and a $c$-antiquark with momentum $p_2^\mu = m_c v'^\mu$, and its  Dirac projector is given by $\mathcal{P}_{\eta} \propto \gamma_5 (1+\slashed{v}')$ with the four-velocity $v'^\mu$ of the $\eta_c$ meson ($v'^2 = 1$). In this approximation, the meson masses $m_{B}$ and $m_{\eta}$, as well as their momenta $p_{B}^\mu$ and $p_{\eta}^\mu$, are simply the sum of the corresponding quark masses and momenta, respectively.

In our analysis, it will be convenient to formally distinguish the masses of the \mbox{``active''} quark that is generated in the flavour-changing weak transition, and the spectator antiquark. We therefore write $p_1^\mu = m_1 v'^\mu$ and $p_2^\mu = m_2 v'^\mu$, along with $\ell^\mu = m_2 v^\mu$, and we will make use of the mass ratios $u_0 = m_1/m_{\eta}$ and $\bar{u}_0 = 1-u_0 = m_2/m_{\eta}$ in the following. Note that in the generalized power counting \mbox{$m_b \gg m_1 \approx m_2 \gg \Lambda_{\rm QCD}$}, these mass ratios are $\mathcal{O}(1)$ quantities, and the physical $B_c \to \eta_c$ case is recovered by setting $u_0 = \bar{u}_0 = 1/2$. 

In the specific kinematic region of interest, the recoiling $\eta_c$ meson has a parametrically large boost $\gamma \equiv v\cdot v' = \mathcal{O}(m_b/m_{\eta})$ in the $B_c$ rest frame. The QCD dynamics is then naturally analysed using light-cone coordinates. Specifically, one introduces two light-like reference vectors $n^\mu$ and $\bar{n}^\mu$ with $n^2 = \bar{n}^2 = 0$ and $n \cdot \bar{n} = 2$ such that any four-vector $k^\mu$ can be decomposed as 
\begin{align}
	k^\mu = k_-\frac{n^\mu}{2} + k_+\frac{\bar{n}^\mu}{2} + k_\perp^\mu \,, 
\end{align}
with $k_- = \bar n \cdot k$, $k_+ = n \cdot k$ and a transverse component $k_\perp^\mu$ that satisfies $k_{\perp}\cdot n =k_{\perp}\cdot \bar n=0$. In this notation, the four-velocities of the incoming $B_c$ and the outgoing $\eta_c$ can be parametrised by $v^\mu = \frac12 n^\mu + \frac12 \bar{n}^\mu$ and $v'^\mu \approx \gamma n^\mu + \frac{1}{4\gamma} \bar{n}^\mu$, respectively. 

While the hadronic matrix elements in $B_c\! \to \eta_c$ transitions can generally be parametrised by three independent form factors, they can be related at leading power in the heavy-quark expansion up to a calculable factorizable term~\cite{Charles:1998dr,Beneke:2000wa}. In the language of SCET, this is usually addressed in a two-step matching procedure, in which one first integrates out hard-momentum fluctuations with virtualities of $\mathcal{O}(m_b^2)$, which leads to an intermediate effective theory called \mbox{SCET-1}. In a second step, which results in the final low-energy effective theory \mbox{SCET-2}, one then integrates out hard-collinear fluctuations with virtualities of $\mathcal{O}(m_b m_{\eta})$ in our notation. Without going into further details here (cf.~e.g.~\cite{Beneke:2015wfa} for a review), one finds that two SCET-1 operators give a leading-power contribution, one of which yielding a factorizable (non-universal) piece that is expressed in terms of meson light-cone distribution amplitudes that are convoluted with a perturbative hard-scattering kernel. The second SCET-1 operator, on the other hand, cannot be factorised in a similar fashion, since the resulting convolutions that involve subleading twist and higher Fock-state contributions would diverge at the endpoints. In the present work, we are precisely interested in this (non-factorizable) ``soft-overlap'' contribution, which can be defined by 
\begin{align}
	\label{eq:soft-overlap}
	F(\gamma) \equiv \frac{1}{2E_\eta} \bra{\eta_c(p_{\eta})} \big( \bar{q}_1 \Gamma b\big)(0) \ket{B_c(p_{B})} \,, \qquad \text{with} \qquad \Gamma = \frac{\slashed{\bar{n}} \slashed{n}}{4} 
\end{align}
and the large energy of the $\eta_c$ meson $E_{\eta} = \gamma m_{\eta}= \mathcal{O}(m_b)$. The form factors are usually considered as functions of the momentum transfer $q^2=(p_{B} - p_{\eta})^2 \approx m_b^2 -2\gamma m_{\eta} m_b$, but we prefer to use the large boost factor $\gamma$ here instead. In fact, as will be discussed in the following sections, the form factor receives double logarithmic corrections $\sim\alpha_s^{n} \ln^{2n} (2\gamma)$ at each order in perturbation theory that we envisage to resum to all orders in this work.

\section{Fixed-order analysis}
\label{sec:fixed-order}

\begin{figure}[t]
	\centering
    \includegraphics[trim = 18 261 18 18,clip,height=0.18\textheight]{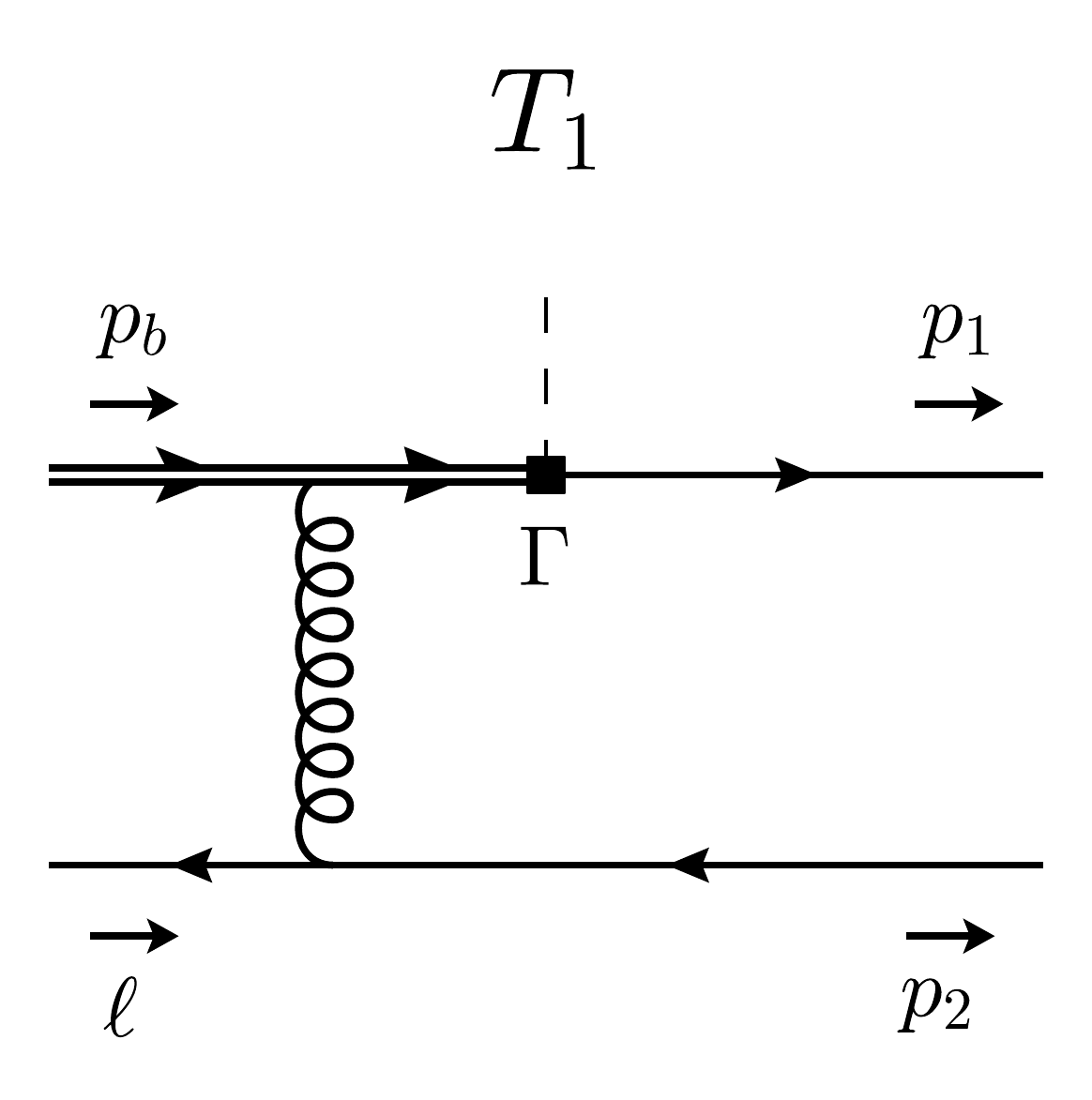} \hspace{1cm}
	\includegraphics[trim = 18 261 18 18,clip,height=0.18\textheight]{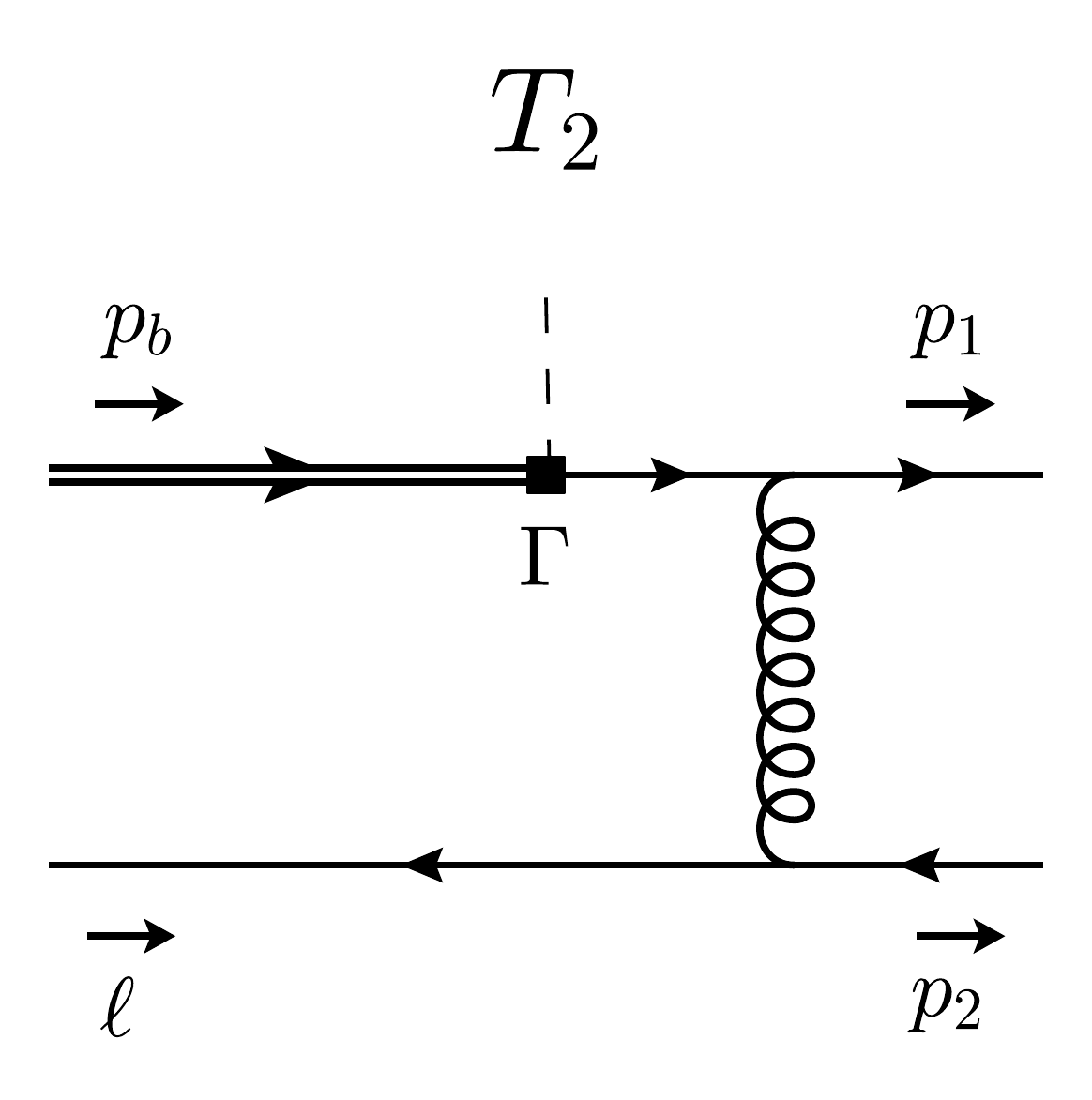}
	\vspace{-0.2cm}
	\caption{Tree-level contributions to the soft-overlap form factor $F(\gamma)$. The double line indicates the incoming heavy $b$-quark, and the remaining solid lines refer to the active light quark with mass $m_1$ (upper line) and spectator antiquark with mass $m_2$ (lower line). We furthermore use $\Gamma = \frac{\slashed{\bar{n}} \slashed{n}}{4}$ and project onto the $B_c$ and $\eta_c$ states using $\mathcal{P}_{B}$ and $\mathcal{P}_{\eta}$ as described in the text. 
	\label{fig:LO:diagrams}}
\end{figure}

At tree level, the form factor receives contributions from the two Feynman diagrams shown in Fig.~\ref{fig:LO:diagrams}. Writing $F(\gamma) = \sum_{n=0}^\infty \left(\frac{\alpha_s}{4\pi}\right)^n F^{(n)}(\gamma) + \mathcal{O}(m_{\eta}/m_b)$, the explicit evaluation of these diagrams yields 
\begin{align}
	\label{eq:xiLO}
	F^{(0)}(\gamma) = \xi_0 \,\frac{2+\bar{u}_0}{\bar{u}_0^3} \,, \qquad \text{with} \qquad \xi_0 = \frac{\alpha_s C_F}{4\pi} \frac{\pi^2 f_{B} f_{\eta} m_B}{N_c E_{\eta}^2 m_{\eta}} \,, 
\end{align}
where we restored the prefactors in the meson projectors,
\begin{align}
	\mathcal{P}_{B} = \frac{i f_{B} m_{B}}{4N_c} \,(1+ \slashed{v})\gamma_5\,,\qquad\quad
	\mathcal{P}_{\eta} =\frac{i f_{\eta} m_{\eta}}{4N_c} \,\gamma_5(1+\slashed{v}')\,, 
\end{align}
which are fixed by the requirement that the matrix element $\bra{0} \bar{q}_2 \gamma^\mu \gamma_5 b \ket{B_c(p_{B})} = i f_{B} p_{B}^\mu$ reduces to the standard definition of the decay constant, and similarly for the $\eta_c$ meson. We also note that we do not distinguish between the QCD and the (scale-dependent) HQET decay constant of the $B_c$ meson here, since their difference is only a single-logarithmic effect, which is beyond the approximation of interest.

The tree-level analysis allows us to highlight another important aspect of the calculation, i.e.~it simplifies considerably in light-cone gauge $\bar{n} \cdot A = 0$. In this gauge, the free gluon propagator takes the form
\begin{align}
	\Delta^{\mu\nu}(k) = 
	\frac{i}{k^2} \left[-g^{\mu\nu}+\frac{\bar{n}^\mu k^\nu+\bar{n}^\nu k^\mu}{\bar{n}\cdot k} \right] 
	= 
	\frac{i}{k^2} \left[-g_\perp^{\mu\nu}+\frac{n\cdot k}{\bar{n}\cdot k} \;\bar{n}^\mu \bar{n}^\nu
	+\frac{\bar{n}^\mu k_\perp^\nu+\bar{n}^\nu k_\perp^\mu}{\bar{n}\cdot k} \right], 
\end{align}
and one easily verifies that the first diagram in Fig.~\ref{fig:LO:diagrams} does not provide a leading-power contribution in this case. Physically, the reason is that energetic gluons couple to the heavy $b$-quark via eikonal Wilson-line interactions,
\begin{align}
	\frac{\Gamma (\!\slashed{~p_b}+\slashed{\ell}-\!\!\slashed{~p_2}+m_b)\gamma^\mu \mathcal{P}_{B}}{(p_b+\ell-p_2)^2-m_b^2}\, 
	\,\approx\, \frac{\Gamma v^\mu \mathcal{P}_{B}}{-v\cdot p_2}\, , 
\end{align}
because of the specific projection properties of the soft-overlap form factor encoded in $\Gamma$. In a contraction with the leading (first) term of the gluon propagator in light-cone gauge $\propto g_\perp^{\mu\nu}$, this term vanishes identically, and as a result the tree-level contribution gets confined to a single diagram in this gauge. This observation will play an important role for the analysis of higher-order corrections below. In the remainder of this section, we examine how the double-logarithmic contributions are generated at next-to-leading order (NLO) and next-to-next-to-leading order (NNLO) in perturbation theory.

\subsection{Next-to-leading order}
\label{subsec:NLO}

From now on we will use the symbol $\simeq$ to indicate the double-logarithmic approximation. A straight-forward evaluation of the one-loop diagrams gives~\cite{Bell:2006tz,Boer:2018mgl}
\begin{align}
	\label{eq:NLO:DL}
	F^{(1)}(\gamma)  \simeq \xi_0 L^2 \left( C_F \frac{1+2\bar{u}_0}{\bar{u}_0^3} - \frac{C_A}{2\bar{u}_0^3}\right) \,, 
	\qquad \text{with} \qquad L \equiv \ln(2\gamma) \,, 
\end{align}
which reveals that the double logarithms contain an Abelian and a non-Abelian component, and they do not trivially factorize to the Born result \eqref{eq:xiLO}, as reflected by the different dependence on the quark-mass ratio $\bar u_0$. It turns out that the double logarithms are generated via two different mechanisms that we dub \emph{soft-quark} and \emph{soft-gluon corrections}. We now address each of these contributions in turn.

\begin{figure}[t]
	\centering
    \includegraphics[trim = 18 334 18 18,clip,height=0.18\textheight]{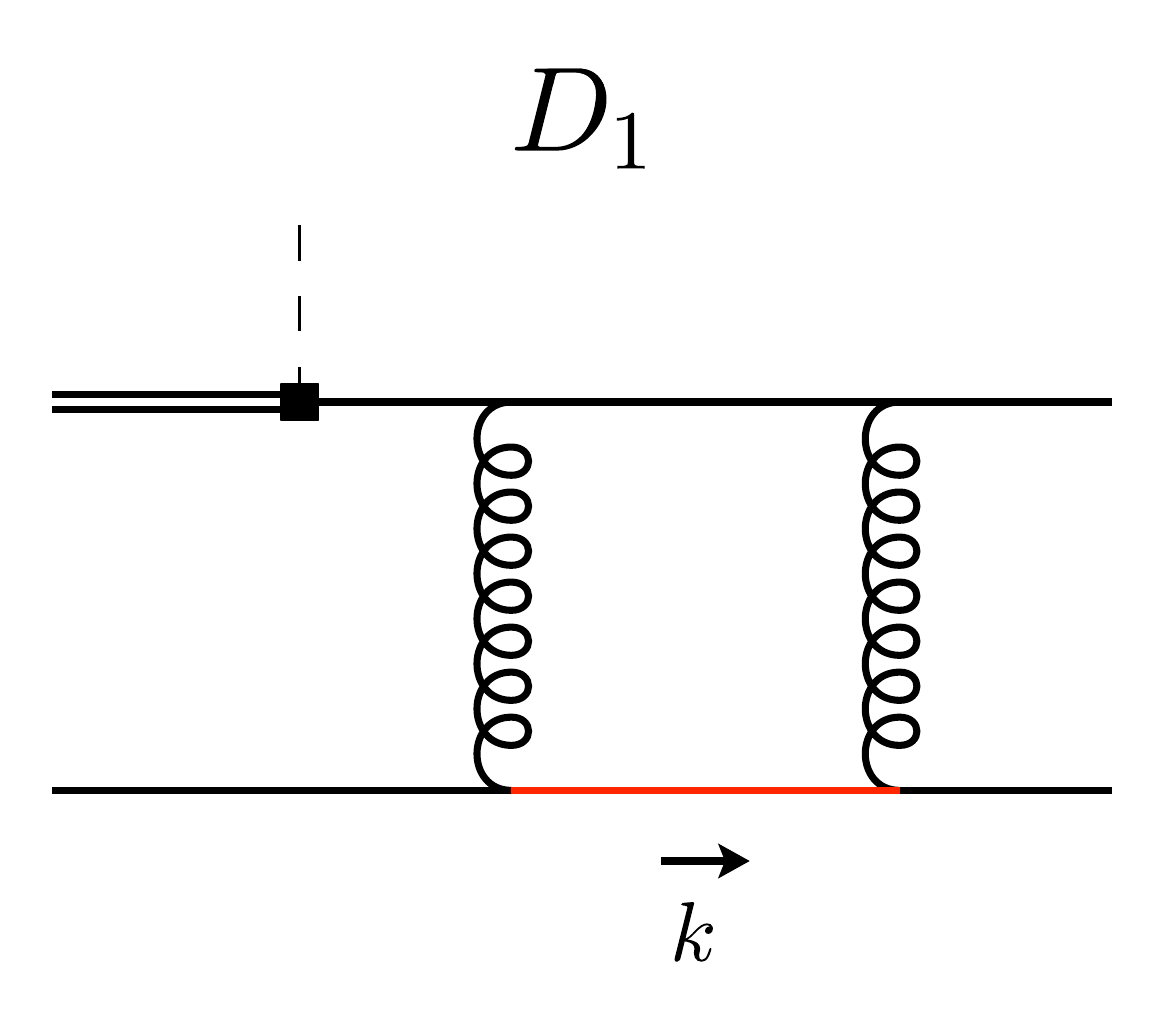} \qquad
	\includegraphics[trim = 18 334 18 18,clip,height=0.18\textheight]{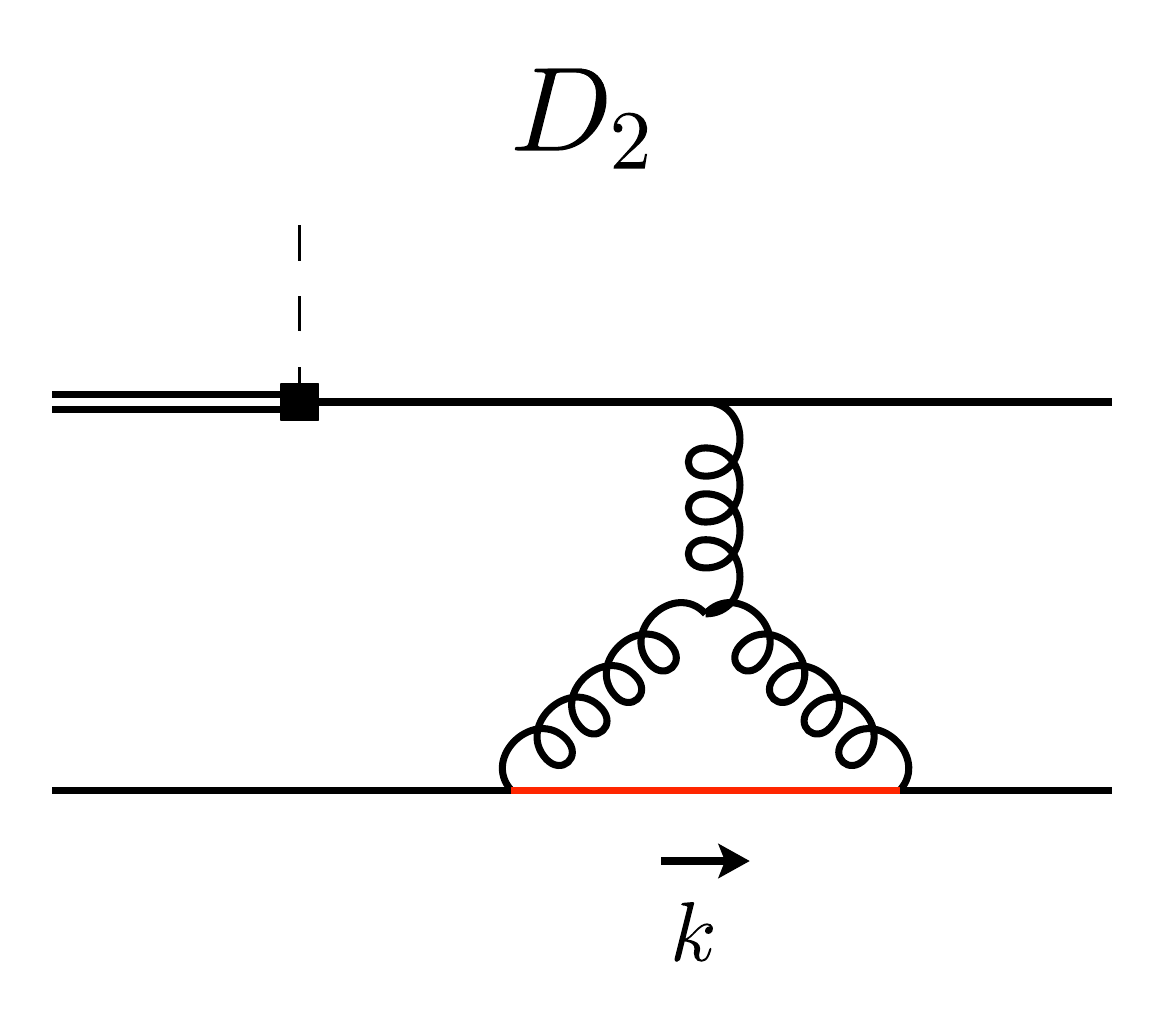} \qquad 
	\raisebox{-0.10em}{\includegraphics[trim = 18 334 18 18,clip,height=0.18\textheight]{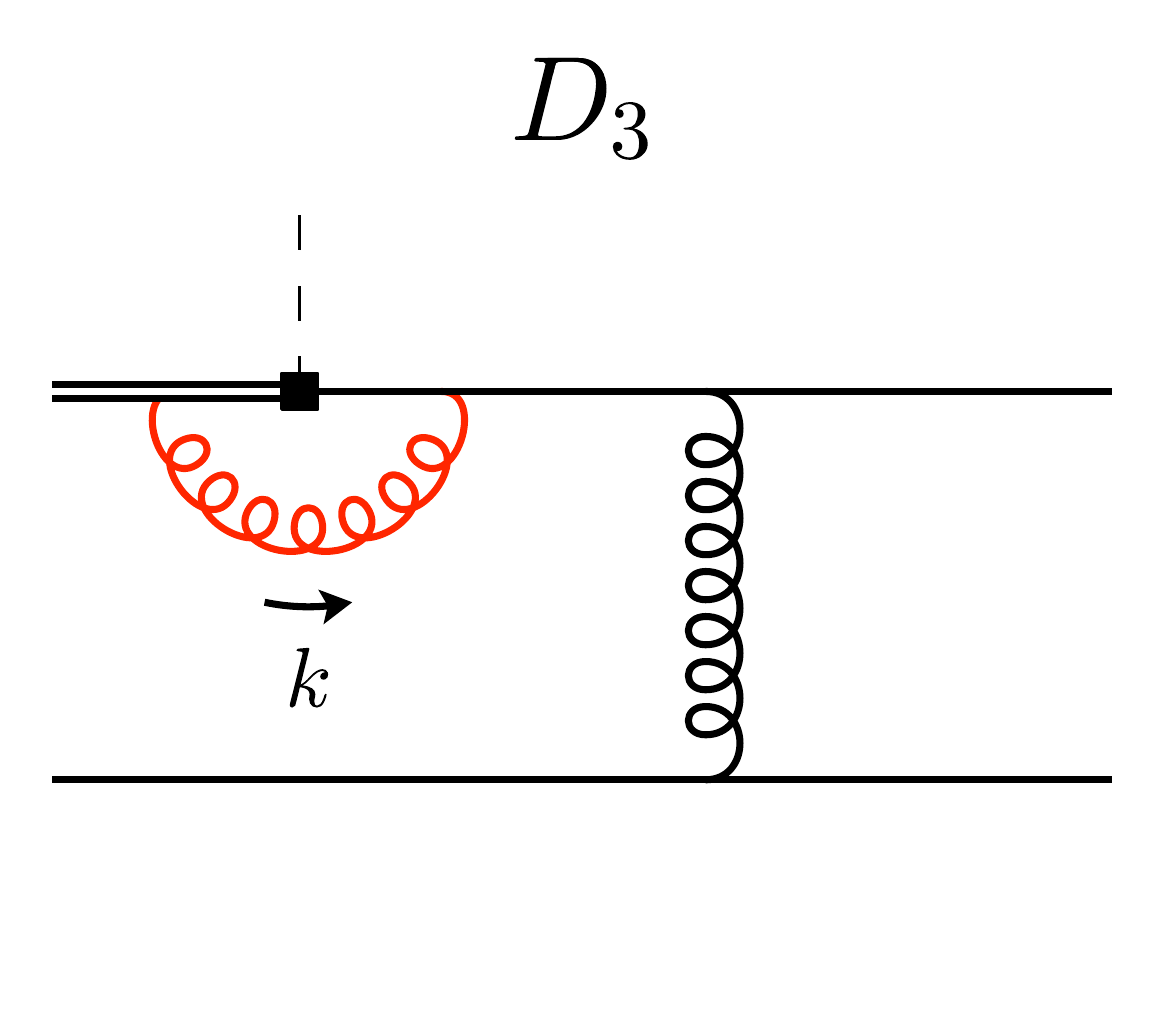}}
	\caption{One-loop contributions that generate a double-logarithmic enhancement. A solid red line indicates a soft-quark and a curly red line a soft-gluon configuration. In light-cone gauge $\bar{n} \cdot A = 0$, the soft-quark corrections are entirely encoded in the first two diagrams, whereas in a hybrid light-cone and Feynman gauge (as described in the text), the soft-gluon corrections can be extracted from the third diagram. 
		\label{fig:NLO:diagrams}}
\end{figure}

\paragraph{Soft-quark corrections:}
First of all, we remark that the terminology may be slightly misleading, since in the language of SCET, one usually refers to the momentum modes in the heavy $B_c$ meson as ``soft'' and to the ones in the energetic light meson as ``collinear''. The configurations we consider here have the same virtuality, but they lie in between these modes in the rapidity ordering, and hence the terminology ``soft-collinear'' could appear more appropriate. We nevertheless prefer to use the notion ``soft'' here, since it connects to the traditional terminology that is used in the literature (see e.g.~\cite{Liu:2017vkm,Liu:2018czl,Bell:2022ott}). 

A convenient method for extracting logarithmic contributions of a loop diagram is the method of regions~\cite{Beneke:1997zp}. In this approach, the logarithms arise from an interplay of distinct dynamical contributions (``momentum regions''), in which the logarithms are usually tied to some sort of divergences. This is, however, not the most efficient method for our purposes, since each individual region by itself already has the information on the double logarithmic terms. We therefore follow a different approach here, in which one extracts the double logarithms from a single region, and imposes \emph{physical cutoffs} to render the contribution well-defined. Specifically, one finds that the soft-quark corrections arise from two one-loop diagrams in light-cone gauge. Starting with the Abelian box diagram that is shown in the left panel of Fig.~\ref{fig:NLO:diagrams}, one has 
\begin{equation}
	D_1 = \int \!\! \frac{d^4k}{(2\pi)^4} \, \frac{\text{Numerator}[D_1]}{[k^2 - m_2^2] \, (\ell-k)^2 \, (k-p_2)^2 \, [(p_\eta -k)^2 - m_1^2] \, [(p_\eta -\ell)^2 - m_1^2]} \,, 
\end{equation}
where the standard $i\epsilon$-prescription of the propagators is under\-stood. We also introduced a symbolic notation for the numerator that represents the Dirac trace that is contracted with the polarization sums of the gluon propagators in light-cone gauge. In the short-hand notation $k^\mu\sim(k_-,k_\perp,k_+)$, in which the external momenta scale as $\ell^\mu\sim(\lambda^2,\lambda^2,\lambda^2)m_b$ and $p_i^\mu\sim(1,\lambda^2,\lambda^4)m_b$ for $\lambda^2=m_\eta/m_b\ll 1$, the relevant momentum region obeys the scaling $k^\mu\sim(\lambda,\lambda^2,\lambda^3)m_b$, which as argued above has the same virtuality of $\mathcal{O}(m_\eta^2)$ as $\ell^\mu$ and $p_i^\mu$, but its rapidity measure $k_-/k_+=\mathcal{O}(1/\lambda^2)$ lies in between $\ell_-/\ell_+=\mathcal{O}(1)$ and $p_{i-}/p_{i+}=\mathcal{O}(1/\lambda^4)$. Keeping only the leading-power terms in this region, one obtains
\begin{equation}
	\label{eq:NLO:D1:integrand}
	D_1 \approx \int \!\! \frac{d^4k}{(2\pi)^4} \, \frac{\text{Numerator}[D_1]}{[k^2 - m_2^2] \, (-\ell_+ k_-) \, (-k_+ p_{2-})\, (-k_+ p_{\eta-}) \, (-\ell_+p_{\eta-} )} \,, 
\end{equation}
which has the remarkable property that all denominators except for the first one associated with the antiquark become eikonal. In standard applications of the method of regions, one would impose suitable (but artificial) regulators to make this contribution well-defined. If one chooses dimensional regularization in combination with a certain analytic rapidity regulator in the spirit of \cite{Becher:2011dz} for this purpose, the above contribution would correspond to a scaleless integral and vanish. Here, in contrast, we isolate the double-logarithmic contribution by imposing physical cutoffs with \mbox{$\ell_-<k_-<p_{2-}$} and $\ell_+>k_+>p_{2+}$ that are fixed by the external kinematics of the process, and by simultaneously constraining the soft quark to be on-shell. In this case, the integral can be directly evaluated in four space-time dimensions without any further regularization.  Specifically, the integration over the transverse components can effectively be performed by replacing (see e.g.~\cite{Liu:2018czl,Bell:2022ott})
\begin{equation}
	\label{eq:on-shell}
	\frac{1}{[k^2 - m_2^2]} \quad\to\quad -2 \pi i \delta(k_+ k_- + k_\perp^2 - m_2^2) \,, 
\end{equation}
and the double-logarithmic sensitivity then arises from the remaining longitudinal integrations that are of the form
\begin{align}
	\label{eq:nested:one-loop}
	\int_{\ell_-}^{p_{2-}} \! \frac{dk_-}{k_-} \int_{p_{2+}}^{\ell_+} \! \frac{dk_+}{k_+} \; 
	\theta(k_+ k_- - m_2^2) \,=  \, \frac12 \ln^2 \bigg(\frac{\ell_+ p_{2-}}{m_2^2}\bigg)
	= \frac12 L^2\,, 
\end{align}
where the $\theta$-function reflects the on-shell condition \eqref{eq:on-shell}. The most non-trivial part of the calculation therefore consists in extracting the relevant terms in the numerator that provide the necessary factor of $k_+$ to bring the integral \eqref{eq:NLO:D1:integrand} into the form \eqref{eq:nested:one-loop} for a double-logarithmic enhancement.  We find, after using the on-shell condition,
\begin{equation}
	\text{Numerator}[D_1] \,\simeq\, 
	384 i \pi \alpha_s C_F \gamma^4 m_\eta^6  (1 + \bar{u}_0) k_+ \,\xi_0\,, 
\end{equation}
which, when assembled with the remaining factors of the diagram, yields a contribution to the soft-overlap form factor that is given by
\begin{equation}
	\label{eq:NLO:D1:result}
	F^{(1)}_{D_1}(\gamma)  \simeq 
	\xi_0 L^2 \left( 3C_F \,\frac{1+\bar{u}_0}{\bar{u}_0^3}  \right). 
\end{equation}
The evaluation of the non-Abelian diagram in the second panel of Fig.~\ref{fig:NLO:diagrams} proceeds similarly. In particular, in the considered momentum region one now has
\begin{equation}
	\label{eq:NLO:D2:integrand}
	D_2 \approx \int \!\! \frac{d^4k}{(2\pi)^4} \, \frac{\text{Numerator}[D_2]}{[k^2 - m_2^2] \, (-\ell_+ k_-) \, (-k_+ p_{2-})\, (-\ell_+ p_{2-}) \, (-\ell_+p_{\eta-} )} \,, 
\end{equation}
where all propagators, except for the one related to the spectator antiquark, again become eikonal. Setting the latter on-shell and invoking the mechanism from \eqref{eq:nested:one-loop} to generate the double-logarithmic sensitivity now requires us to identify the terms in the corresponding numerator that are independent of $k_+$ and $k_-$, since their powers in \eqref{eq:NLO:D2:integrand} already match the ones required by \eqref{eq:nested:one-loop}. With
\begin{equation}
	\text{Numerator}[D_2] \,\simeq\, 
	-64 i \pi \alpha_s C_A \gamma^4 m_\eta^7  \bar{u}_0^2  \,\xi_0\,, 
\end{equation}
the contribution of the non-Abelian diagram then becomes
\begin{equation}
		\label{eq:NLO:D2:result}
	F^{(1)}_{D_2}(\gamma)  \simeq 
	\xi_0 L^2 \left( -\frac{C_A}{2\bar{u}_0^3}  \right). 
\end{equation}
To summarize, the soft-quark corrections are associated with momentum configurations in which the spectator antiquark is on-shell and its longitudinal momentum components are strongly ordered between the adjacent momenta, which ensures that all remaining propagators become eikonal. Given that energetic gluons decouple from the heavy quark at leading power in light-cone gauge, the double logarithms can be clearly localized in one Abelian and one non-Abelian contribution.

\paragraph{Soft-gluon corrections:} The computation of the second class of corrections is rather standard, and will be discussed only briefly here. In fact, as the soft-gluon corrections constitute a gauge-invariant subset, we may choose a specific gauge -- in this case Feynman gauge -- to localize these corrections in a minimal number of relevant Feynman diagrams. As the soft-gluon corrections factorize to the Born amplitude, one may even choose different gauges for the Born (hard-collinear) and dressed (soft) gluons, which from a QCD perspective may seem unusual. In an effective-theory interpretation, on the other hand, this is a very natural procedure, since the structure of gauge transformations is richer in the effective theory than in full QCD (see e.g.~\cite{Becher:2014oda}). We will therefore choose such a hybrid light-cone (for hard-collinear gluons) and Feynman (for soft gluons) gauge in the following.

While there are several diagrams that generate double logarithmic corrections from soft-gluon exchanges in this setup, many of these cancel when the appropriate attachments to the outgoing quark and antiquark are summed up. The total soft-gluon contribution can then be extracted from a single diagram, which is shown in the third panel of Fig.~\ref{fig:NLO:diagrams}. Although one could easily compute this diagram with standard method-of-regions techniques, we follow a different strategy here, which is analogous to the one we used for the soft-quark corrections above. Specifically, one may isolate the double logarithms in a particular region with scaling $k^\mu\sim(\lambda^{1/2},\lambda,\lambda^{3/2})m_b$, which has the same virtuality as a hard-collinear momentum $q_{hc}^\mu\sim(1,\lambda,\lambda^{2})m_b$, but is once again shifted in rapidity. We note that the terminology ``soft'' is again ambiguous here, but the key feature is that all propagators, except for the one of the gluon itself, become eikonal in this region.\footnote{We note that this overlap region would yield a scaleless integral when the standard method-of-regions technique based on dimensional regularisation (and without any cutoffs) is used.} One then immediately sees that the contribution factorizes to the Born amplitude with
\begin{equation}
	\label{eq:NLO:D3:integrand}
	D_3 \approx \, T_2 \times 4\pi\alpha_s C_F \int \!\! \frac{d^4k}{(2\pi)^4} \, \frac{(-2i m_b p_{\eta-})}{[k^2] \, (-m_b \, k_-)  (-k_+ p_{\eta-})} \,, 
\end{equation}
which has a similar structure as the integrals in \eqref{eq:NLO:D1:integrand} and \eqref{eq:NLO:D2:integrand} above, except that the numerator is trivial in this case, reflecting the spin-independent eikonal soft-gluon couplings. Setting the gluon on-shell, $1/k^2 \to -2 \pi i \delta(k_+ k_- + k_\perp^2)$, one then obtains the following integral that encodes the double-logarithmic sensitivity of the soft-gluon corrections at one loop,
\begin{align}
	\label{eq:Sudakov:one-loop}
	\int_{\ell_+}^{p_{\eta-}} \! \frac{dk_-}{k_-} \int_{\ell_+}^{k_-} \! \frac{dk_+}{k_+}  
	\,=  \, \frac12 \ln^2 \bigg(\frac{p_{\eta-}}{\ell_+}\bigg)
	\,=  \, \frac12 \ln^2 \bigg(\frac{2\gamma}{\bar{u}_0}\bigg)
	\simeq \frac12 L^2\,.  
\end{align}
Here the UV cutoff $p_{\eta-}$ and the IR cutoff $\ell_{+}$ arise from the phase-space boundaries of the (real) soft-gluon-emission process. In particular, $p_{\eta-}=\mathcal{O}(m_b)$ reflects the maximal energy that can be emitted from the heavy quark, and $\ell_{+}=\mathcal{O}(m_\eta)$ is related to the virtuality of the adjacent hard-collinear line with momentum $(p_\eta - \ell)$. The remaining boundary $k_+<k_-$ is critical to generate the correct double-logarithmic coefficient, and its effect is to turn a time-like Wilson line that would be aligned along the $v^\mu$ direction into a light-like Wilson line along $\bar{n}^\mu$. Restoring the prefactors of the diagram, one then finds that the soft-gluon corrections to the soft-overlap form factor become
\begin{equation}
	\label{eq:NLO:D3:result}	
	F^{(1)}_{D_3}(\gamma)  \simeq 
	\xi_0 L^2 \left( -C_F \,\frac{2+\bar{u}_0}{\bar{u}_0^3}  \right) = -C_F L^2 \,F^{(0)}(\gamma)\,. 
\end{equation}
Adding the soft-quark corrections from \eqref{eq:NLO:D1:result} and \eqref{eq:NLO:D2:result} to this expression then correctly reproduces the result \eqref{eq:NLO:DL} of the one-loop computation.

\subsection{Next-to-next-to-leading order}
\label{subsec:NNLO}

Having identified the dynamical origin of the double-logarithmic enhancement at one loop, we will now examine how this picture extends to the next order. Specifically, we will see that the dominant logarithms originate from double soft-quark, double soft-gluon and mixed soft-quark and soft-gluon insertions. We will now analyze these contributions in turn.

\begin{figure}[t]
	\centering
	\includegraphics[trim = 18 371 18 18,clip,height=0.18\textheight]{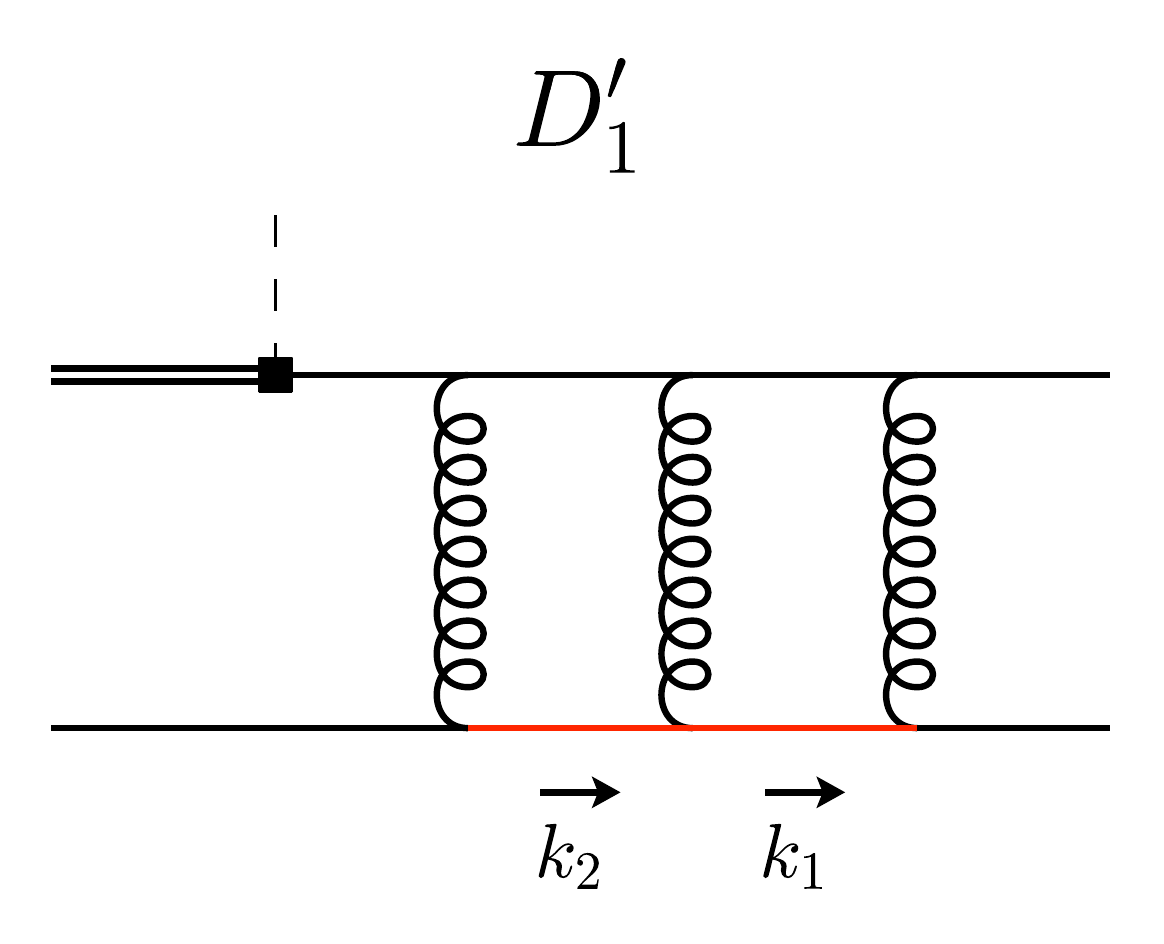} 
	\includegraphics[trim = 18 371 18 18,clip,height=0.18\textheight]{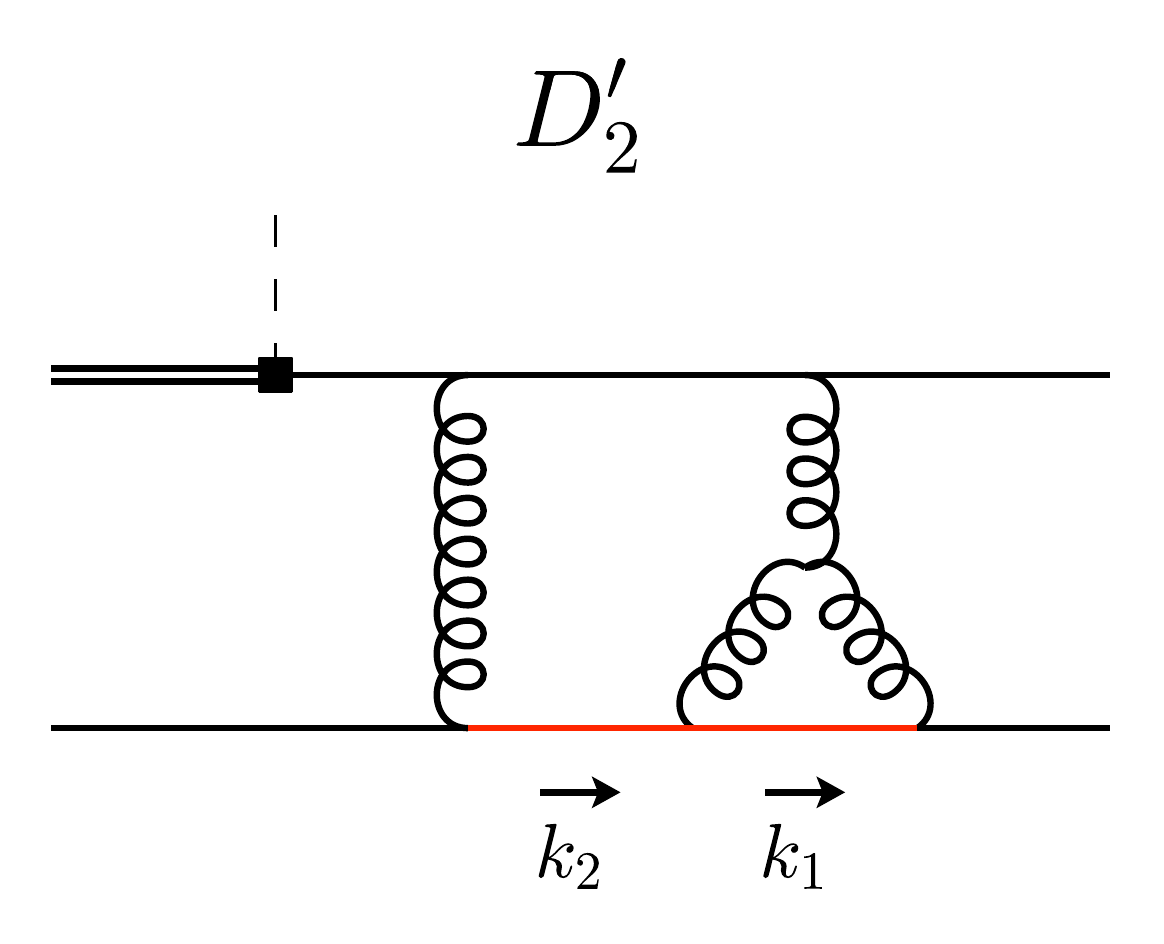} 
	\includegraphics[trim = 18 371 18 18, clip,height=0.18\textheight]{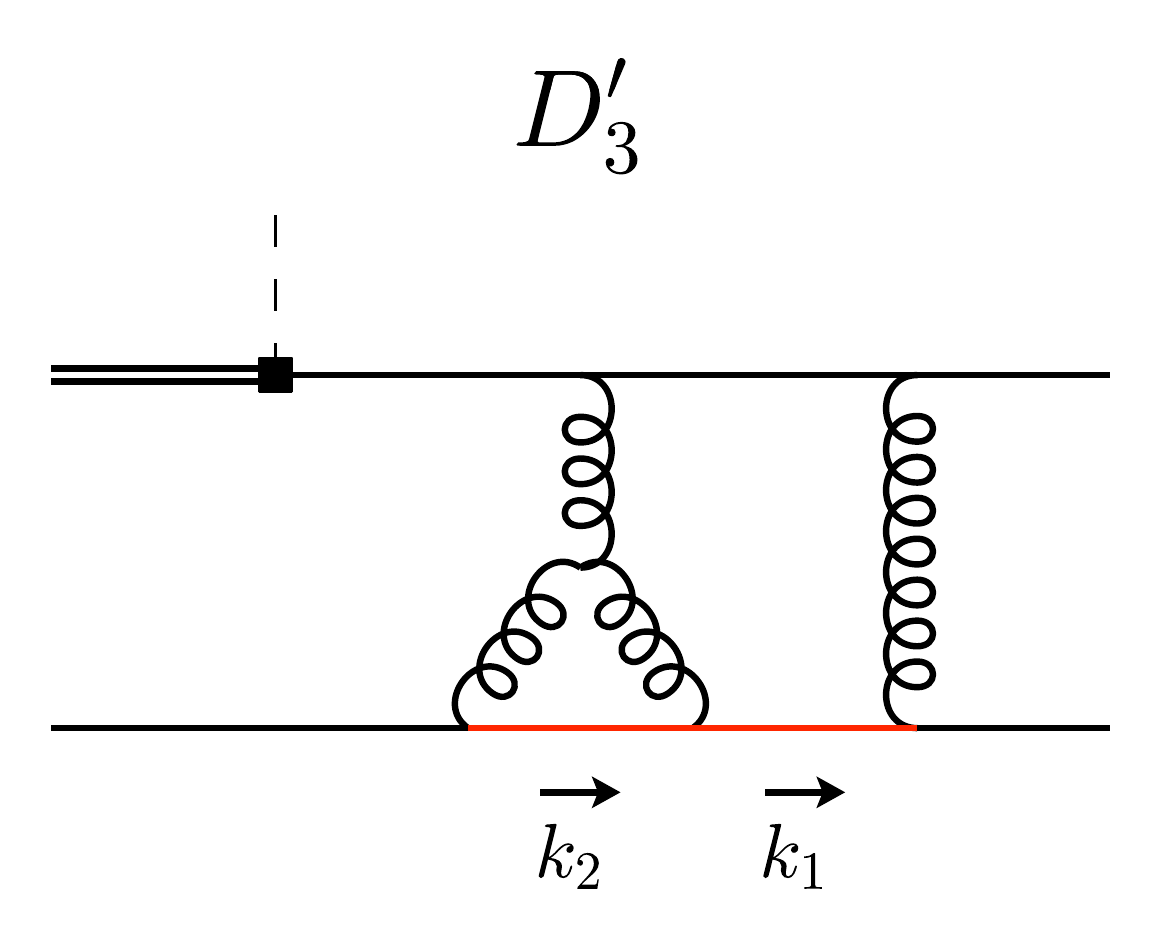}
	\vspace{-0.2cm}
	\caption{Two-loop diagrams that capture the leading logarithmic corrections from double soft-quark insertions in light-cone gauge. }
	\label{fig:NNLO:double-soft-quark:diagrams}
\end{figure}

\paragraph{Double soft-quark corrections:}
In light-cone gauge, we find that this class of corrections is captured by the three diagrams shown in Fig.~\ref{fig:NNLO:double-soft-quark:diagrams}, where the loop momenta $k_1^\mu$ and $k_2^\mu$ are assigned from right to left as illustrated in the figure. Technically, the logarithms can be extracted from the region with $k_1^\mu\sim(\lambda^{2/3},\lambda^2,\lambda^{10/3})m_b$ and $k_2^\mu\sim(\lambda^{4/3},\lambda^2,\lambda^{8/3})m_b$, i.e.~the loop momenta satisfy the \emph{strong-ordering prescription} $\ell_-<k_{2-} <k_{1-} <p_{2-}$ and $\ell_+>k_{2+}>k_{1+}>p_{2+}$, whereas their virtuality is again of $\mathcal{O}(m_\eta^2)$. As before, we impose the on-shell condition~\eqref{eq:on-shell} for the spectator-quark propagators, whereas all remaining propagators become eikonal in this region. After integrating over the transverse momentum components, the logarithmic sensitivity then arises from the \emph{nested integrals}
\begin{align}
	\label{eq:nested:two-loop}
	\int_{\ell_-}^{p_{2-}} \! \frac{dk_{2-}}{k_{2-}} 
	\int_{k_{2-}}^{p_{2-}} \! \frac{dk_{1-}}{k_{1-}} 
	\int_{m_2^2/k_{2-}}^{\ell_+} \! \frac{dk_{2+}}{k_{2+}} 
	\int_{m_2^2/k_{1-}}^{k_{2+}} \! \frac{dk_{1+}}{k_{1+}} 
	\,=  \, \frac{1}{12} L^4\,. 
\end{align}
This is similar to the mechanism relevant for energetic muon-electron backward scattering described in~\cite{Gorshkov:1966qd,Bell:2022ott}. Focusing for concreteness on the Abelian box diagram, the calculation thus starts from the representation
\begin{equation}
	\label{eq:NNLO:D1p:intermediate}
	D_1^\prime \approx \int \!\! \frac{d^4k_1}{(2\pi)^4} \frac{d^4k_2}{(2\pi)^4}\, 
	\frac{\text{Numerator}[D_1^\prime]}{[k_1^2 - m_2^2]  [k_2^2 - m_2^2]  (\ell_+ k_{2-})  (k_{2+} k_{1-}) (k_{1+} p_{2-}) (k_{1+} p_{\eta-})  (k_{2+} p_{\eta-})  (\ell_+p_{\eta-} )} \,, 
\end{equation}
and by counting the various plus- and minus-components in the denominator, one concludes that the numerator must provide a factor of $k_{1+} k_{2+}$ to generate the leading logarithmic enhancement according to \eqref{eq:nested:two-loop}. While we postpone a systematic analysis of these numerator structures to the following section, here we quote only the result,
\begin{equation}
	\text{Numerator}[D_1^\prime] \,\simeq\, 
	- 8192 \pi^2 \alpha_s^2 C_F^2 \gamma^5 m_\eta^7  (1 + \bar{u}_0) k_{1+}k_{2+} \,\xi_0\,, 
\end{equation}
which translates into the following contribution to the soft-overlap form factor,
\begin{equation}
	\label{eq:NNLO:D1p:result}
	F^{(2)}_{D_1^\prime}(\gamma)  \simeq 
	\xi_0 L^4 \left( \frac43 C_F^2 \,\frac{1+\bar{u}_0}{\bar{u}_0^3}  \right). 
\end{equation}
The remaining two diagrams can be evaluated along similar lines, and we do not go into further details here. They evaluate to
\begin{equation}
	\label{eq:NNLO:D2pD3p:result}
	F^{(2)}_{D_2^\prime}(\gamma)  \simeq 
	\xi_0 L^4 \left( - \frac{C_A C_F}{6\bar{u}_0^3}  \right),
	\qquad\quad
	F^{(2)}_{D_3^\prime}(\gamma)  \simeq 
	\xi_0 L^4 \left( - C_A C_F \,\frac{1+\bar{u}_0}{6\bar{u}_0^3}  \right), 
\end{equation}
which results in the following total double soft-quark contribution
\begin{equation}
	\label{eq:NNLO:soft-quark:result}	
	F^{(2)}_{\text{soft quarks}}(\gamma)  \simeq 
	\xi_0 L^4 \left( \frac43 C_F^2 \,\frac{1+\bar{u}_0}{\bar{u}_0^3}  
    - C_A C_F \,\frac{2+\bar{u}_0}{6\bar{u}_0^3} \right). 
\end{equation}

\paragraph{Double soft-gluon corrections:}
At NLO we saw that the structure of the soft-gluon corrections is much simpler, since they trivially factorize to the Born amplitude because of the spin-independent nature of the soft-gluon couplings in the eikonal approximation. Working as before in a hybrid light-cone and Feynman gauge, the task then consists in dressing the right tree-level diagram of Fig.~\ref{fig:LO:diagrams} with two soft gluons. The result is, of course, well known: The one-loop soft-gluon correction exponentiates to all orders. The total double soft-gluon contribution to the form factor therefore amounts to
\begin{equation}
	\label{eq:NNLO:soft-gluon:result}	
	F^{(2)}_{\text{soft gluons}}(\gamma)  \simeq 
	\xi_0 L^4 \left( \frac12 C_F^2 \,\frac{2+\bar{u}_0}{\bar{u}_0^3}  \right) = \frac12 C_F^2 L^4 \,F^{(0)}(\gamma) \, . 
\end{equation}

\begin{figure}[p!]
	\centering
    \newlength{\myplotwidthPS}
    \newlength{\myplotwidthF}
    \setlength{\myplotwidthPS}{0.27\textwidth}
    \setlength{\myplotwidthF}{0.45\textwidth}
    \includegraphics[trim = 18 543 18 18,clip,width=\myplotwidthF]{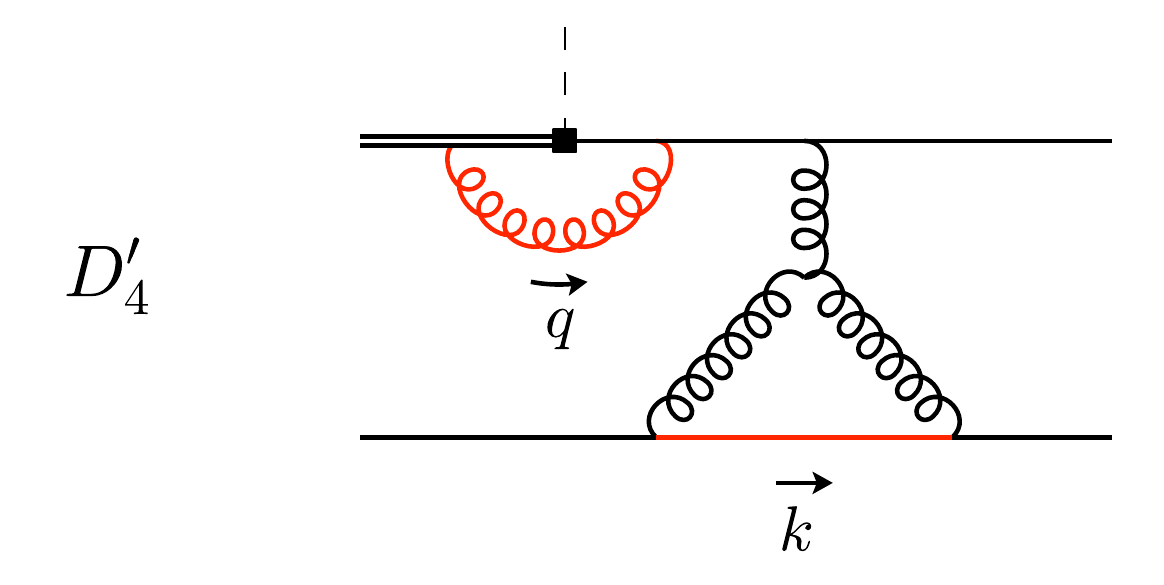}
	   \hspace{1.8cm}
    \raisebox{0.95em}{\includegraphics[width=\myplotwidthPS]{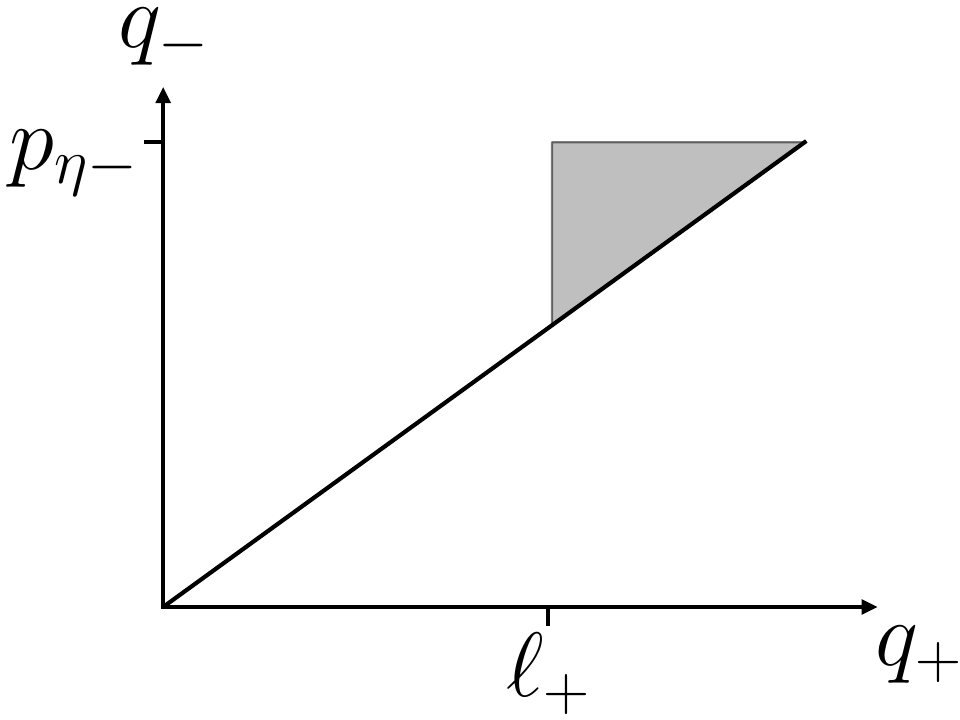}}\\[0.7em]
	\includegraphics[trim = 18 543 18 18,clip,width=\myplotwidthF]{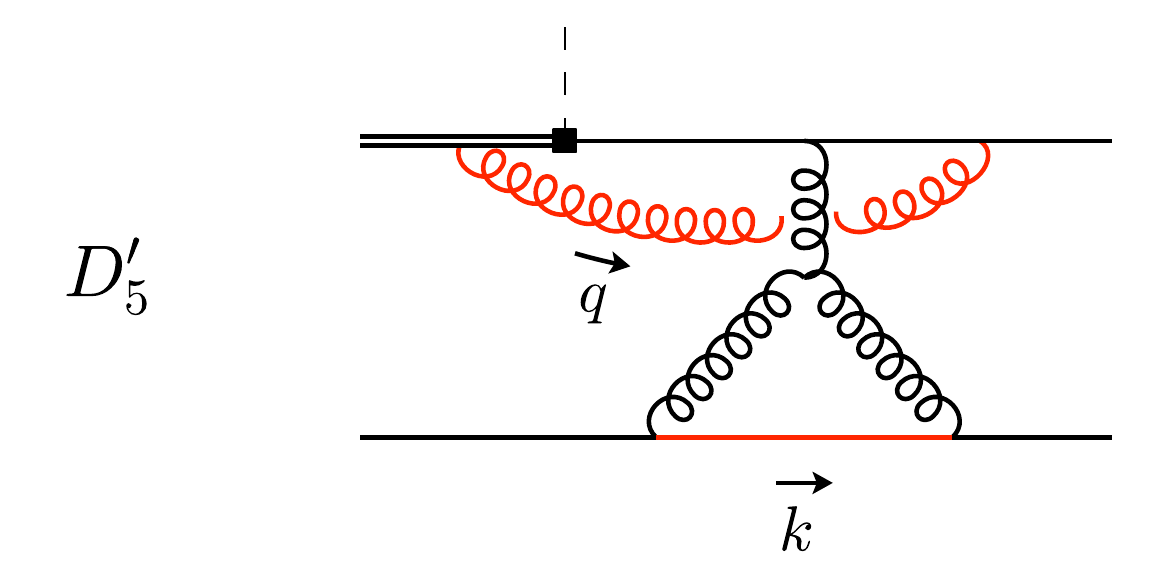} \hspace{1.8cm}
    \raisebox{0.95em}{\includegraphics[width=\myplotwidthPS]{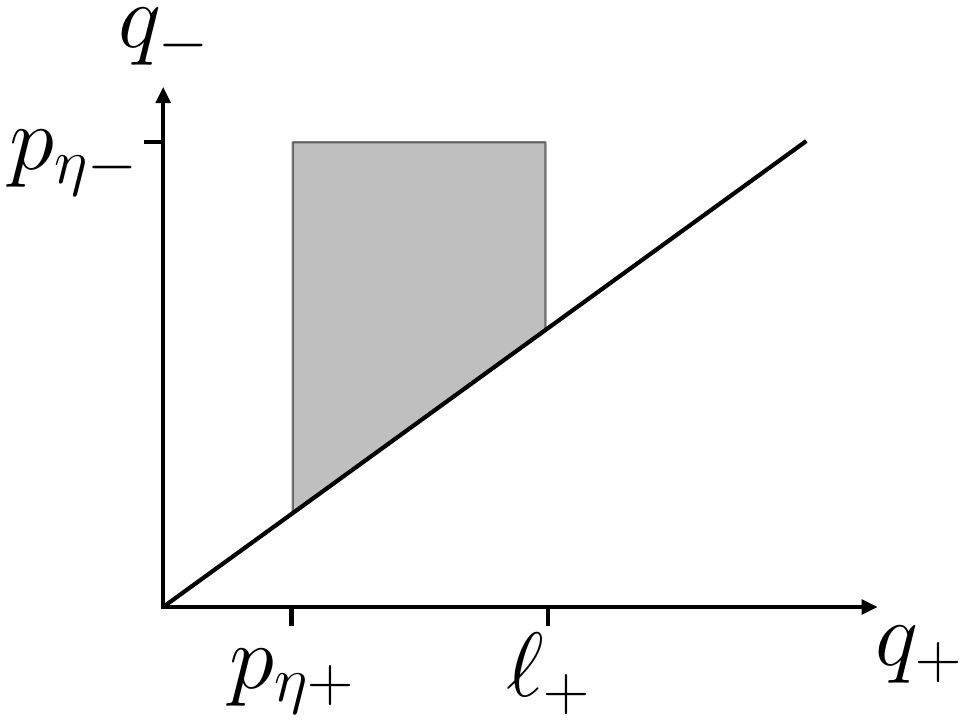}}\\[0.7em]
	\includegraphics[trim = 18 543 18 18,clip,width=\myplotwidthF]{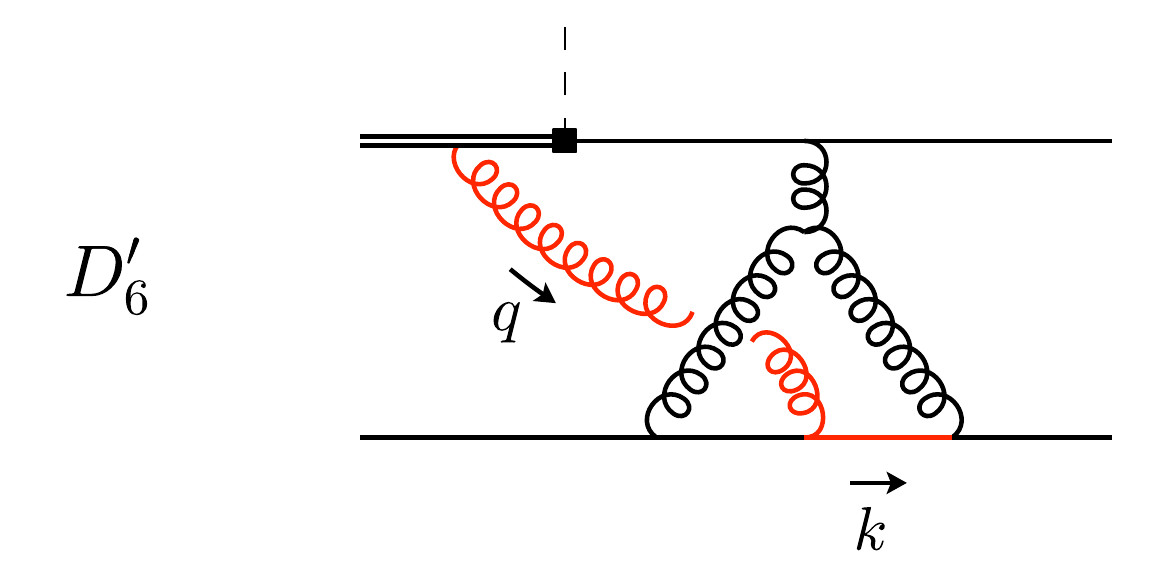}
    \hspace{1.8cm}
    \raisebox{0.95em}{\includegraphics[width=\myplotwidthPS]{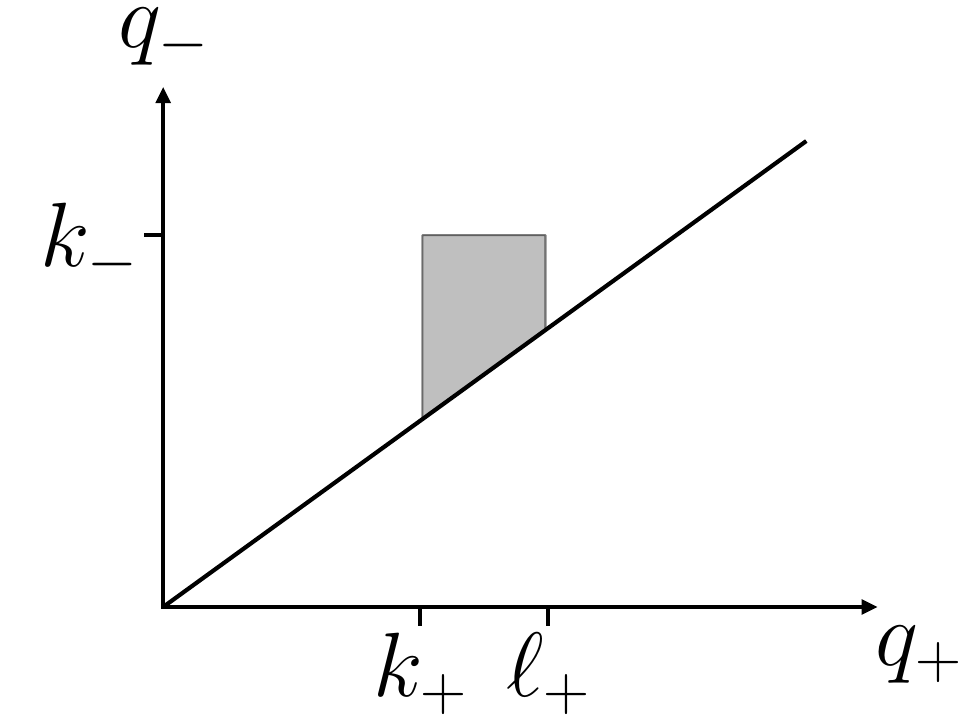}}\\[0.7em]
	\includegraphics[trim = 18 543 18 18,clip,width=\myplotwidthF]{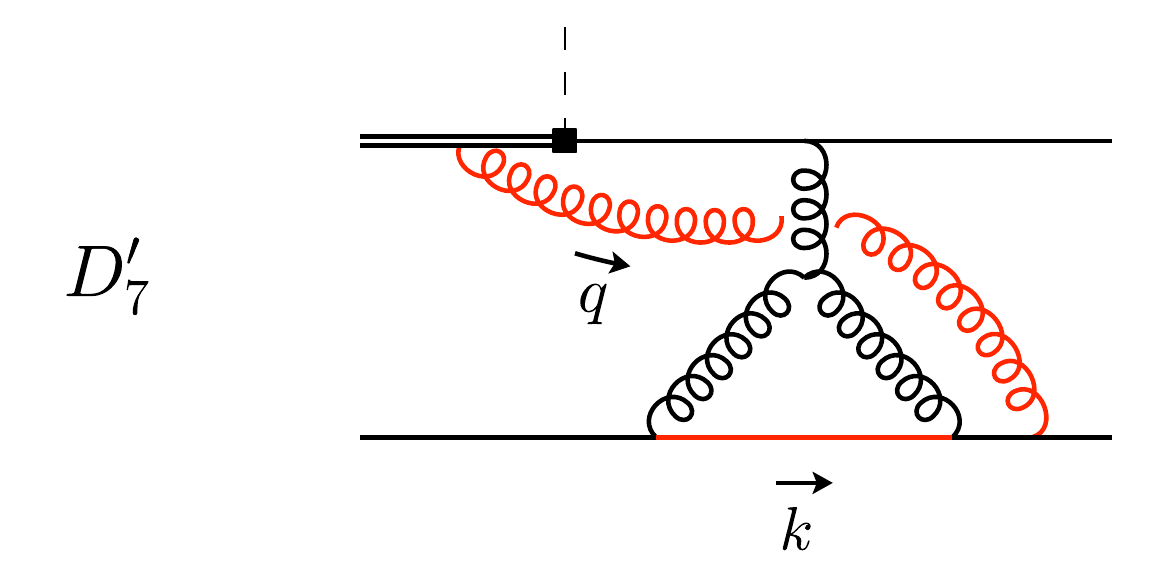}
    \hspace{1.8cm}
    \raisebox{0.95em}{\includegraphics[width=\myplotwidthPS]{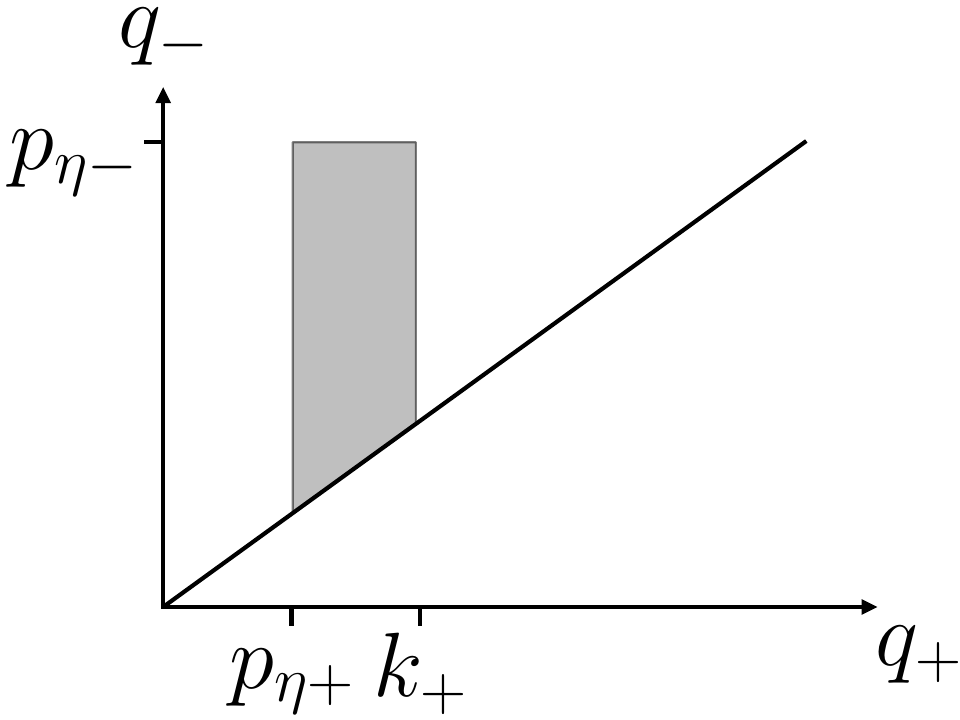}}\\[0.7em]
	\includegraphics[trim = 18 543 18 18,clip,width=\myplotwidthF]{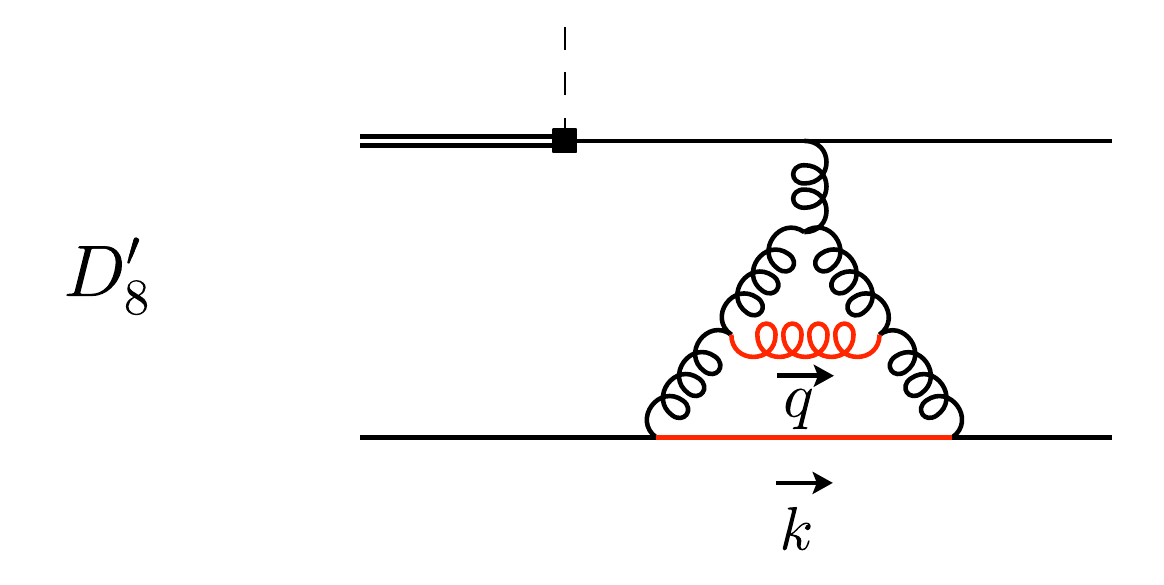}
    \hspace{1.8cm}
    \raisebox{0.95em}{\includegraphics[width=\myplotwidthPS]{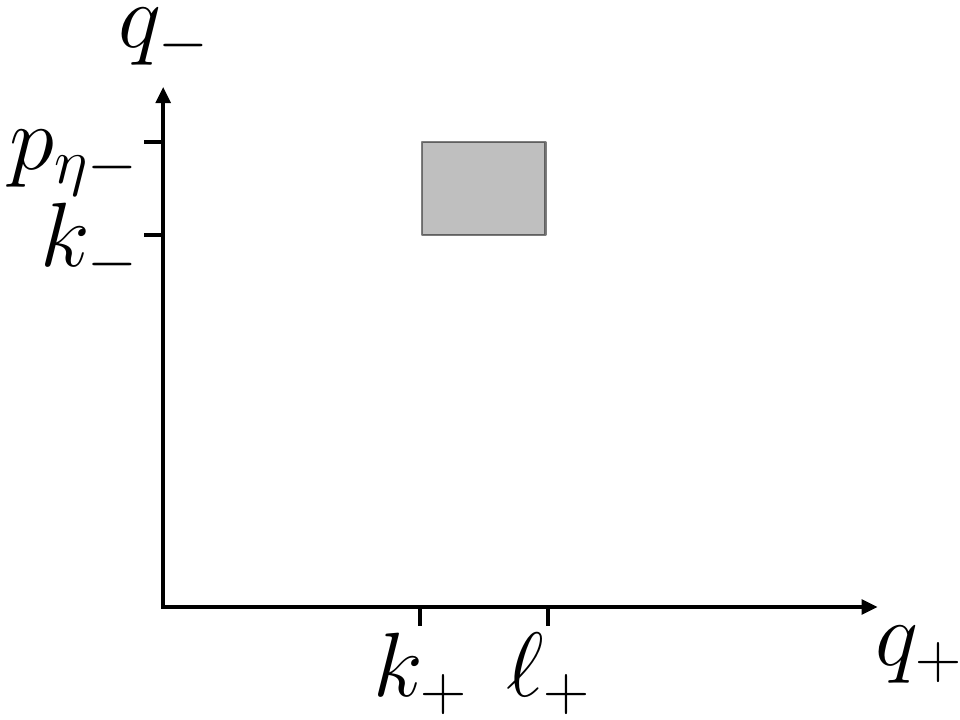}}
	\vspace{-0.0cm}
	\caption{Subset of two-loop diagrams that yield leading logarithmic corrections from the interplay of soft-quark and soft-gluon configurations in a hybrid light-cone and Feynman gauge. For each diagram we display the available phase space for the soft-gluon emission.}
	\label{fig:NNLO:interplay:diagrams}
\end{figure}

\paragraph{Mixed soft-quark and soft-gluon corrections:}
The interplay of the soft-quark and soft-gluon corrections, which for the first time arises at this order, is more interesting. In the hybrid gauge, these effects can be  extracted from those diagrams that dress the one-loop soft-quark configurations in Fig.~\ref{fig:NLO:diagrams} with all sorts of soft-gluon attachments. Starting, for concreteness, with the non-Abelian diagram $D_2$, we find that the dominant logarithms are encoded in the five diagrams shown in Fig.~\ref{fig:NNLO:interplay:diagrams}. While each of these diagrams trivially factorizes to $D_2$, which is the relevant ``Born'' configuration for the soft-gluon attachments, the key point is that the relevant phase space for the soft-gluon emission with momentum $q^\mu$  differs for each diagram, as illustrated in the figure. These phase-space boundaries can be found by eikonalizing all propagators except for the soft-quark and soft-gluon propagators themselves. Specifically, we find that the first diagram yields a simple multiplicative correction,
\begin{equation}
	\label{eq:NNLO:D4p:result}
	D_4' \simeq \, D_2 \times \frac{\alpha_s}{4\pi} (-C_F) L^2\,, 
\end{equation}
which is the same overall factor we found at one-loop order in \eqref{eq:NLO:D3:result}. The contribution from the remaining diagrams is, on the other hand, more involved, since the phase-space boundaries for the soft-gluon emission depend on the soft-quark momentum $k^\mu$ in this case. Given that eikonal attachments to a quark and an antiquark contribute with opposite signs, the sum of diagrams $D_5'$, $D_6'$ and $D_7'$ combines into
\begin{equation}
	\label{eq:NNLO:D5pD6pD7p:result}
	D_5' + D_6' + D_7' \simeq \, D_2 \otimes \frac{\alpha_s}{4\pi} \, (-2)
    \Big(C_F - \frac{C_A}{2}\Big) \,\ln \frac{p_{\eta -}}{k_-} \,\ln \frac{\ell_+}{k_+} \,, 
\end{equation}
where the symbol $\otimes$ indicates a convolution, i.e.~the double logarithm from the soft-gluon emission modifies the \emph{integrand} of the soft-quark loop in \eqref{eq:nested:one-loop}. The last diagram in Fig.~\ref{fig:NNLO:interplay:diagrams} then simply cancels the $C_A$-type contribution,
\begin{equation}
	\label{eq:NNLO:D8p:result}
	D_8' \simeq \, D_2 \otimes \frac{\alpha_s}{4\pi} \, (-C_A) \,\ln \frac{p_{\eta -}}{k_-} \,\ln \frac{\ell_+}{k_+} \,, 
\end{equation}
yielding in total a Sudakov-type correction proportional to the Casimir in the fundamental representation. We explicitly verified that the soft-gluon attachments to the box diagram $D_1$ take an analogous form -- a global correction factor as in \eqref{eq:NNLO:D4p:result} and a non-trivial interplay of soft-quark and soft-gluon corrections provided by the sum of  \eqref{eq:NNLO:D5pD6pD7p:result} and \eqref{eq:NNLO:D8p:result}, but now proportional to $D_1$ -- although the detailed diagrammatic analysis turns out to be more complicated in this case. This is related to the fact that the Dirac structure of the NLO box diagram contains contributions from the small components of the hard-collinear quark spinor, which in SCET are integrated out via the equations of motion. As a result the soft-gluon corrections are spread over a larger number of diagrams with more complicated phase-space boundaries.

Adding up the full set of mixed soft-quark and soft-gluon corrections and performing the final convolutions over the soft-quark momentum $k^\mu$ then leads to the following leading-logarithmic contribution,
\begin{equation}
	\label{eq:NNLO:interplay:result}	
	F^{(2)}_{\text{mixed}}(\gamma)  \simeq 
	\xi_0 L^4 \left( 
    -3C_F^2 \,\frac{1+\bar{u}_0}{\bar{u}_0^3}  
     + \frac{C_A C_F}{2\bar{u}_0^3} \right)
     +\xi_0 L^4 \left( 
    -C_F^2 \,\frac{1+\bar{u}_0}{2\bar{u}_0^3}  
     + \frac{C_A C_F}{12\bar{u}_0^3} \right), 
\end{equation}
where the first term arises from the global Sudakov factor in \eqref{eq:NNLO:D4p:result}, and the second term reflects the non-trivial interplay of the soft-quark and soft-gluon corrections.\\

By adding the various contributions from double soft quarks in \eqref{eq:NNLO:soft-quark:result}, double soft gluons in \eqref{eq:NNLO:soft-gluon:result} and the mixed soft-quark and soft-gluon corrections in \eqref{eq:NNLO:interplay:result}, we obtain the total leading logarithmic two-loop correction to the soft-overlap form factor. We find
\begin{equation}
	\label{eq:NNLO:full:result}	
	F^{(2)}(\gamma)  \simeq 
	\xi_0 L^4 \left( 
    -C_F^2 \,\frac{7+10\bar{u}_0}{6\bar{u}_0^3}  
     + C_A C_F\, \frac{3-2\bar{u}_0}{12\bar{u}_0^3} \right), 
\end{equation}
which has a very non-trivial structure with respect to both color and the quark-mass dependence encoded in powers of $\bar{u}_0$, due to the various mechanisms that play together. In order to verify this result, we have performed an independent cross-check based on a bare factorization formula derived in~\cite{Boer:2018mgl}. More specifically, we computed the leading $1/\epsilon^4$ singularities of the two-loop amplitude in the purely hard-collinear momentum region, which can be related to the required coefficient of the leading logarithmic correction using pole-cancellation arguments. For further details of this cross-check, we refer to a future publication~\cite{BBFHS}.

\section{All-order structure}
\label{sec:all-order}

The diagrammatic analysis of the previous section reveals certain patterns that we will now generalize to all orders. Specifically, the nested scalar integrals that appear in the soft-quark corrections are familiar from muon-electron backward scattering~\cite{Gorshkov:1966qd,Bell:2022ott}, but one needs to translate this mechanism into a different kinematic regime in a non-Abelian context. Moreover, this class of integrals has not been studied in the presence of soft-gluon corrections before. We will now address each of these aspects in turn, which will lead to a system of coupled integral equations that capture the double-logarithmic corrections to the soft-overlap form factor to all orders in perturbation theory.

\subsection{Soft-quark corrections}
\label{sec:all-order:soft-quark}

\begin{figure}[t]
    \centering
    \includegraphics[trim = 18 546 18 18,clip,height=0.165\textheight]{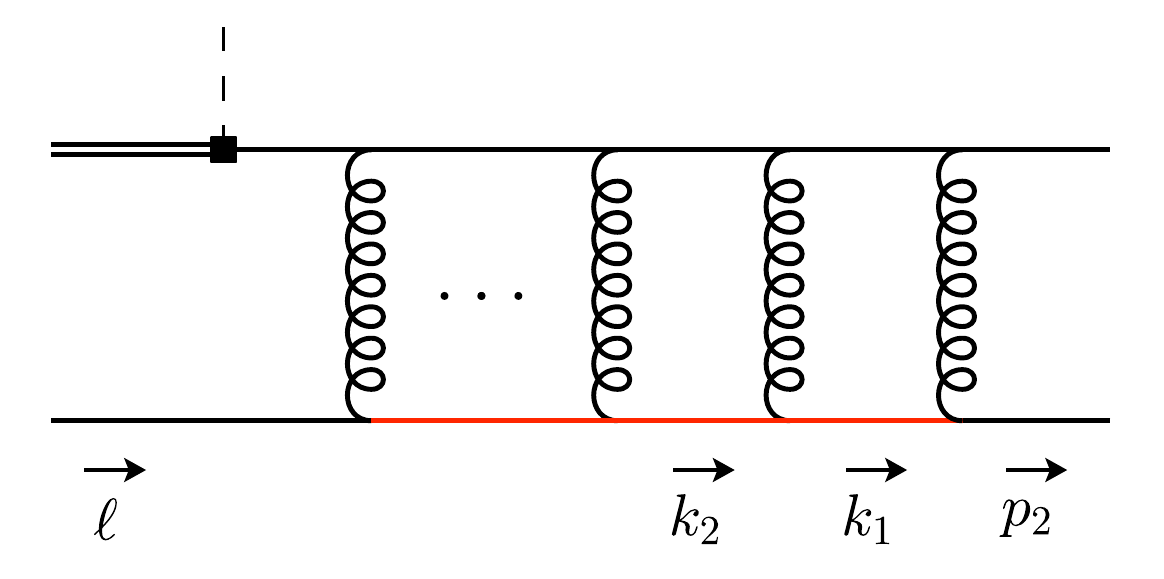} \hspace{1cm}
    \includegraphics[trim = 18 546 18 18,clip,height=0.165\textheight]{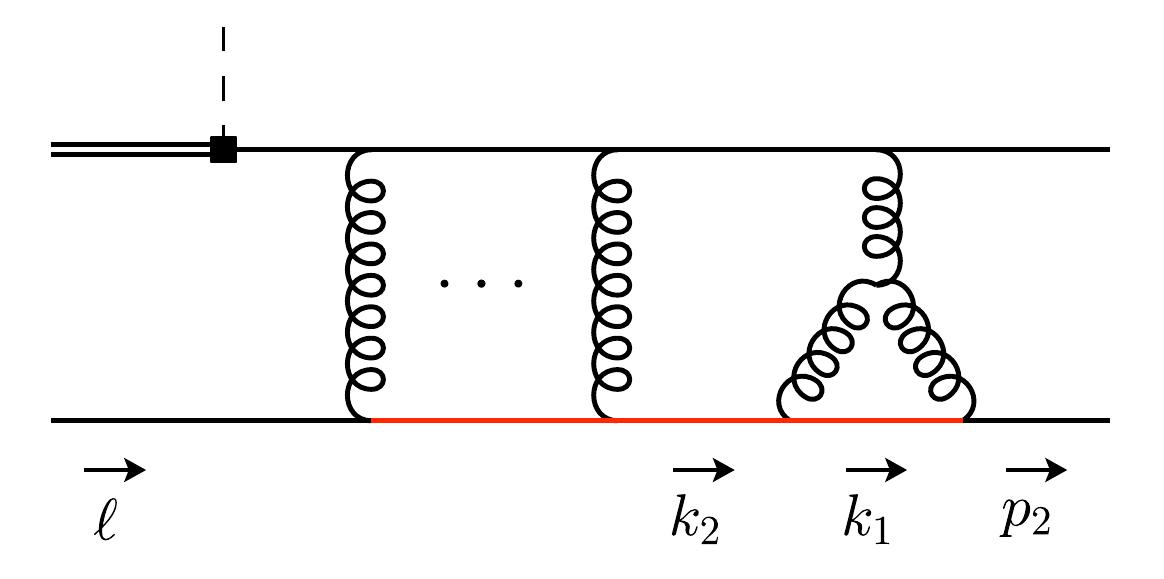}
    \vspace{-0.2cm}
    \caption{Sample diagrams that generate leading logarithmic corrections from soft-quark configurations in light-cone gauge.}
\label{fig:SoftQuarkLadder}
\end{figure}

We start by considering the pure soft-quark corrections, which are captured by two (three) diagrams at NLO (NNLO) in light-cone gauge. At any order in perturbation theory, the double-logarithmic enhancement is associated with configurations in which the spectator antiquark momenta $k_i^\mu$ with \mbox{$i=1,\ldots,n$} are strongly ordered,
\begin{align}
	& \ell_-< k_{n-} < k_{(n-1)-} < \ldots < k_{1-} <p_{2-}\,, 
	\nonumber\\
	& \ell_+>k_{n+}> k_{(n-1)+}> \ldots > k_{1+}>p_{2+}\,, 
	\label{eq:strong-ordering}
\end{align}
with $k_{i-}k_{i+}\approx k_{i\perp}^2=\mathcal{O}(m_\eta^2)$. The specific assignment of the particle momenta is illustrated in Fig.~\ref{fig:SoftQuarkLadder}. In this configuration, all propagators in the relevant Feynman diagrams become eikonal, except for the spectator-antiquark propagators themselves. Setting the latter on-shell then leads to a generalized class of nested integrals of the form
\begin{align}
\label{eq:nested:n-loop}
  & \int_{\ell_-}^{p_{2-}} \! \frac{dk_{n-}}{k_{n-}} 
	\int_{k_{n-}}^{p_{2-}} \! \frac{dk_{(n-1)-}}{k_{(n-1)-}} 
	\; \dots 
	\int_{k_{2-}}^{p_{2-}} \! \frac{dk_{1-}}{k_{1-}}
   \, 
	\int_{m_2^2/k_{n-}}^{\ell_+} \!\! \frac{dk_{n+}}{k_{n+}} 
	\int_{m_2^2/k_{(n-1)-}}^{k_{n+}} \!\! \frac{dk_{(n-1)+}}{k_{(n-1)+}} 	\; \dots 
	\int_{m_2^2/k_{1-}}^{k_{2+}} \!\! \frac{dk_{1+}}{k_{1+}} 
	\nonumber\\
	&\quad
	= \frac{1}{n!(n+1)!} \; L^{2n} \, , 
\end{align}
that capture the double-logarithmic enhancement at any order in perturbation theory. While this class of integrals is familiar from muon-electron backward scattering~\cite{Gorshkov:1966qd,Bell:2022ott}, the key difference here is the non-trivial Dirac structure. As can be read off from \eqref{eq:NLO:D1:integrand} or \eqref{eq:NNLO:D1p:intermediate}, the eikonal factors in the denominators of the loop integrals may not necessarily match the ones required by \eqref{eq:nested:n-loop}, and the numerators must therefore counterbalance the exceeding factors for a leading-logarithmic sensitivity. Understanding the systematics of these numerators is the main objective of this section. To approach this question, we will successively integrate over a single loop momentum at a time, starting from the right ($\eta_c$) side of the relevant Feynman diagrams, i.e.~in increasing order of the loop momenta $k_i^\mu$. In order to keep the analysis transparent, we will first focus on the Abelian ladder diagrams, and we will subsequently add the non-Abelian contributions. We also  verified through explicit computation of potentially relevant diagrams with soft-quark configurations that there are no further topologies that generate double-logarithmic corrections at higher orders.

\paragraph{Ladder diagrams:} We will first build up the diagrams that yield the Abelian soft-quark contribution~\cite{Boer:2023tcs}. Starting with the right-most gluon exchange at tree level, we obtain in light-cone gauge 
\begin{align}
\label{eq:all-order:ladder:initial}
\vcenter{\hbox{\vspace{0.0cm} \includegraphics[trim = 18 41 19 18,clip,height=3cm]{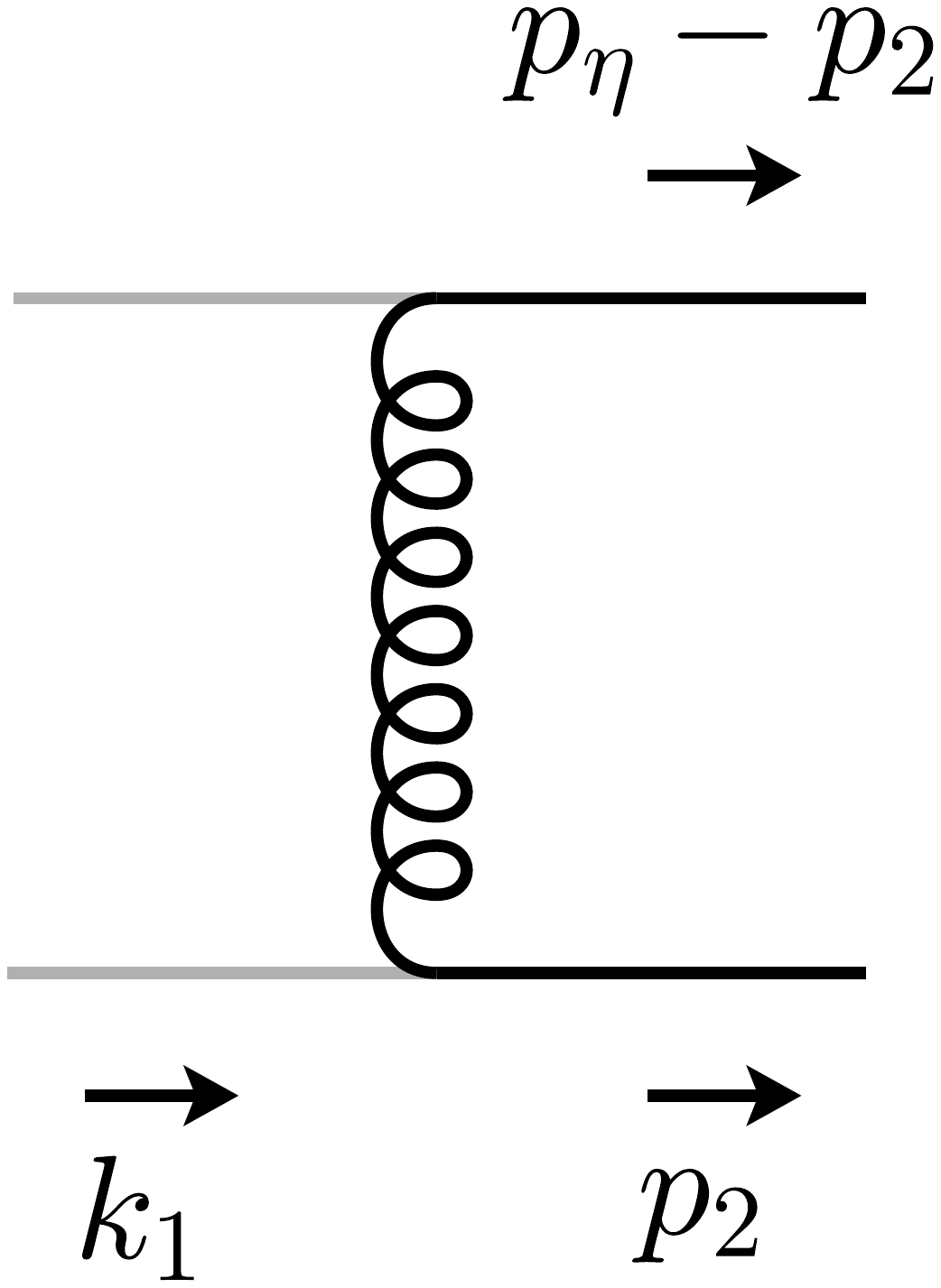}}} \;
&=\; \frac{1}{k_{1+}} \, \sum_{i=1}^7 \, c_i^{(0)} \, \Gamma_i(k_1) \gamma_5 \,, 
\end{align}
where we expressed the result in terms of a basis of Dirac matrices defined by\footnote{We will see later that some of the basis elements do not yield double-logarithmic contributions at leading power, but we find it instructive to consider the general case first.}
\begin{align}
\label{eq:all-order:basis}
	\Gamma_i(k_1) =\bigg\{ \gamma \slashed{n}, 1 , \frac{\slashed{k}_{1,\perp} + k_{1+}\slashed{\bar{n}}}{\bar{u}_0 m_\eta},
	\frac{\slashed{\bar{n}}\slashed{n}}{4}, \frac{\slashed{k}_{1,\perp} \slashed{\bar{n}}}{4\bar{u}_0 \gamma m_\eta}, \frac{\slashed{k}_{1,\perp} \slashed{n}}{2k_{1+}}, \frac{\slashed{k}_{1,\perp} \slashed{\bar{n}}\slashed{n}}{2\bar{u}_0 m_\eta} \bigg\}\,. 
\end{align}
The pictorial representation we use in \eqref{eq:all-order:ladder:initial} and below is to be understood in the sense that black (or red) quantities are included in the calculation, whereas the gray lines anticipate the next (or previous) iteration step in the soft-quark ladder. Specifically in \eqref{eq:all-order:ladder:initial}, one thus has to evaluate a Dirac string with the light-meson projector $\mathcal{P}_{\eta}$ sandwiched between two Dirac matrices from the QCD vertices that are contracted with the gluon propagator in light-cone gauge. The explicit calculation yields
\begin{align}
c_1^{(0)} = c_2^{(0)} = - c_3^{(0)} \approx 
-\frac{\pi \alpha_s C_F f_{\eta}}{N_c \gamma \bar{u}_0}\,,
\qquad\qquad c_4^{(0)}=c_5^{(0)}=c_6^{(0)}=c_7^{(0)}=0 \, , 
\end{align}    
where we kept only the leading-power contribution for each of the coefficients. Adding the next rung to the ladder and assuming the hierarchy \eqref{eq:strong-ordering} for the spectator-antiquark momenta then leads to a contribution of the form		
\begin{align}		
		\vcenter{\hbox{\vspace{+0.0cm} \includegraphics[trim = 18 289 18 18,clip,height=3cm]{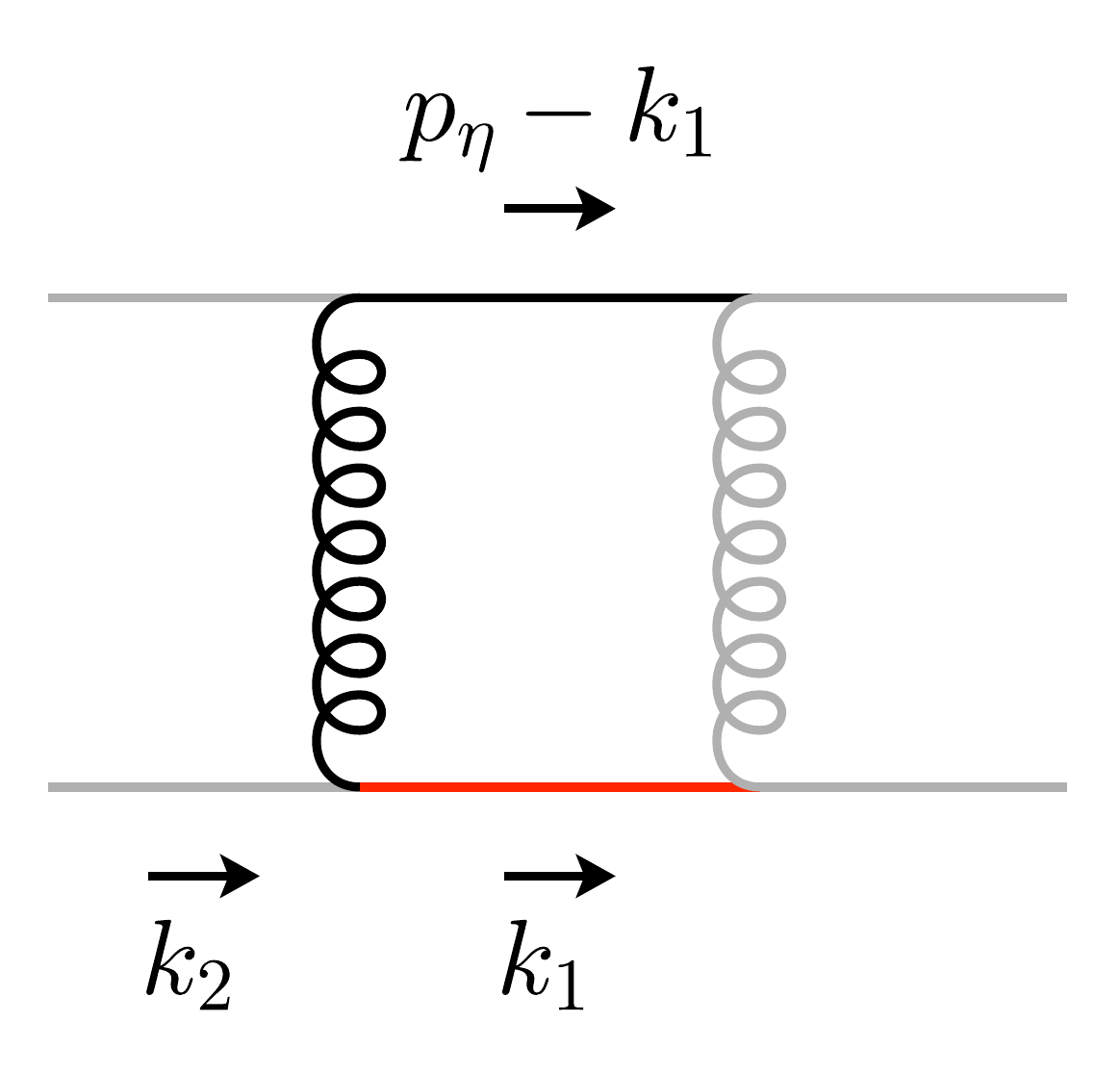}}} \;
\approx\; \int \!\! \frac{d^4k_1}{(2\pi)^4} \; 
\frac{\text{Numerator}}{[k_1^2 - m_2^2] \, (-k_{1+} p_{\eta-}) \, (-k_{2+} k_{1-}) \, k_{1+}} \,, 
\end{align}
from which we read off that the numerator must provide a single factor of $k_{1+}$ and no further factor of $k_{1-}$, after using the on-shell condition \eqref{eq:on-shell}. The result can then again be expressed in the form \eqref{eq:all-order:ladder:initial} with $k_1^\mu \to k_2^\mu$ and a new set of coefficients $c_i^{(1)}$. Explicitly, we find $c_1^{(1)} = c_6^{(1)} = c_7^{(1)} =0$ and\footnote{Notice that the coefficients $c_i^{(n)}$ include factors of  $\alpha_s/(4\pi)$, which is different from the notation we used in the previous section.}
\begin{align}
	\label{eq:all-order:ladder:coefficients}
	\left. \begin{array}{l}
		c_2^{(1)} \\[0.5em]
		c_3^{(1)} \\[0.5em]
		c_4^{(1)} \\[0.5em]
		c_5^{(1)} 
	\end{array} \right\} \simeq
	\frac{\alpha_s C_F}{2\pi} 
	\int_{k_{2-}}^{p_{2-}} \! \frac{dk_{1-}}{k_{1-}}
	\int_{m_2^2/k_{1-}}^{k_{2+}} \!\! \frac{dk_{1+}}{k_{1+}} \;
    \left\{ \begin{array}{l}
	-(1-\bar{u}_0)c_1^{(0)} + c_2^{(0)} - c_3^{(0)}, \\[0.5em]
	- \bar{u}_0 \,c_1^{(0)} + c_3^{(0)}, \\[0.5em]
	c_1^{(0)} - c_2^{(0)}, \\[0.5em]
	c_5^{(0)} - 2 c_7^{(0)}. 
	\end{array} \right. 
\end{align} 
Adding further rungs to the ladder yields precisely the same pattern with a new set of coefficients $c_i^{(j)}$ that are determined by the lower-order coefficients $c_i^{(j-1)}$ via \eqref{eq:all-order:ladder:coefficients}. The iteration then builds up the nested integrals from \eqref{eq:nested:n-loop}, and one finally needs to close the Dirac trace after the last rung of the ladder.
For an $n$-loop diagram (with $n+1$ rungs), this yields 
\begin{align}
\label{eq:all-order:final-projection}
\vcenter{\hbox{\vspace{+0.0cm} \includegraphics[trim = 18 504 18 18,clip,height=3cm]{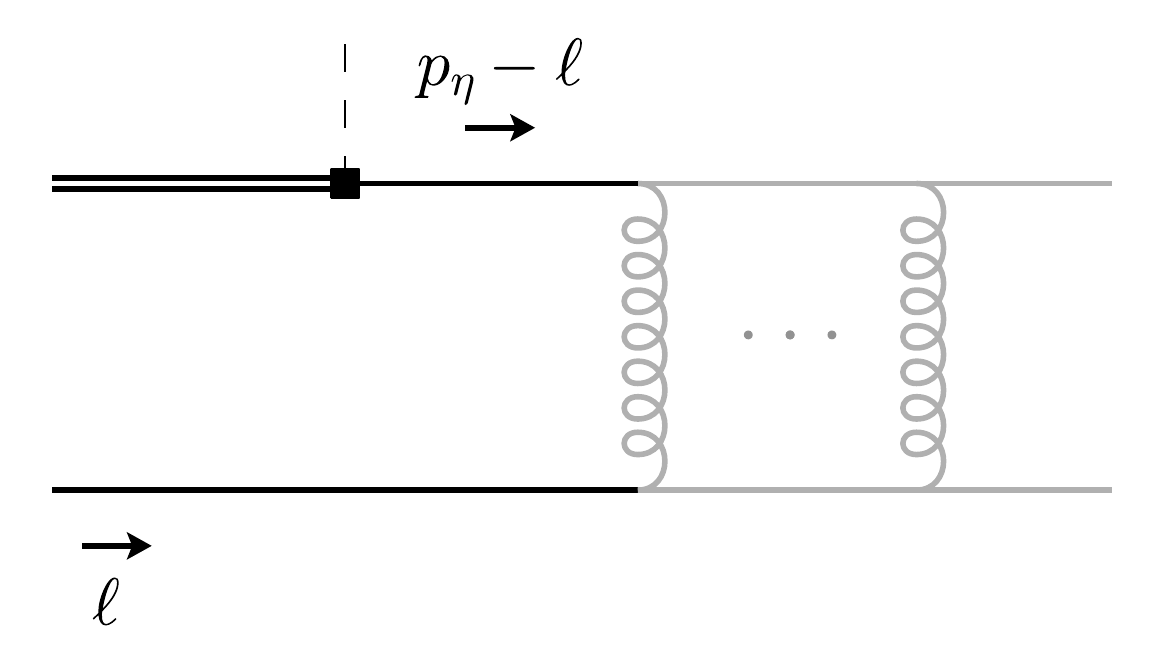}}}  \;\approx\;
\frac{f_B m_B}{2 \bar{u}_0^2 m_\eta^2} \,
\bigg[(1 - \bar{u}_0) c_1^{(n)} - c_2^{(n)} + 2 c_3^{(n)} \bigg]\,. 
\end{align}
We thus see that only the coefficients $c_i^{(n)}$ with $i=1,2,3$ contribute in the final projection, which according to \eqref{eq:all-order:ladder:coefficients} mix among themselves for the considered class of diagrams. This observation significantly simplifies the all-order analysis, as we will see below. One furthermore verifies that the above prescription correctly reproduces the contributions from the tree-level, one-loop and two-loop ladder diagrams given in \eqref{eq:xiLO}, \eqref{eq:NLO:D1:result} and \eqref{eq:NNLO:D1p:result}, respectively. 

\paragraph{Non-Abelian contributions:} 
We next consider diagrams with particular insertions of three-gluon vertices. Specifically, for the right-most insertion, we obtain in the hierarchy \eqref{eq:strong-ordering}  of the spectator-antiquark momenta
\begin{align}		
	\vcenter{\hbox{\vspace{+0.0cm} \includegraphics[trim = 18 276 18 18,clip,height=3cm]{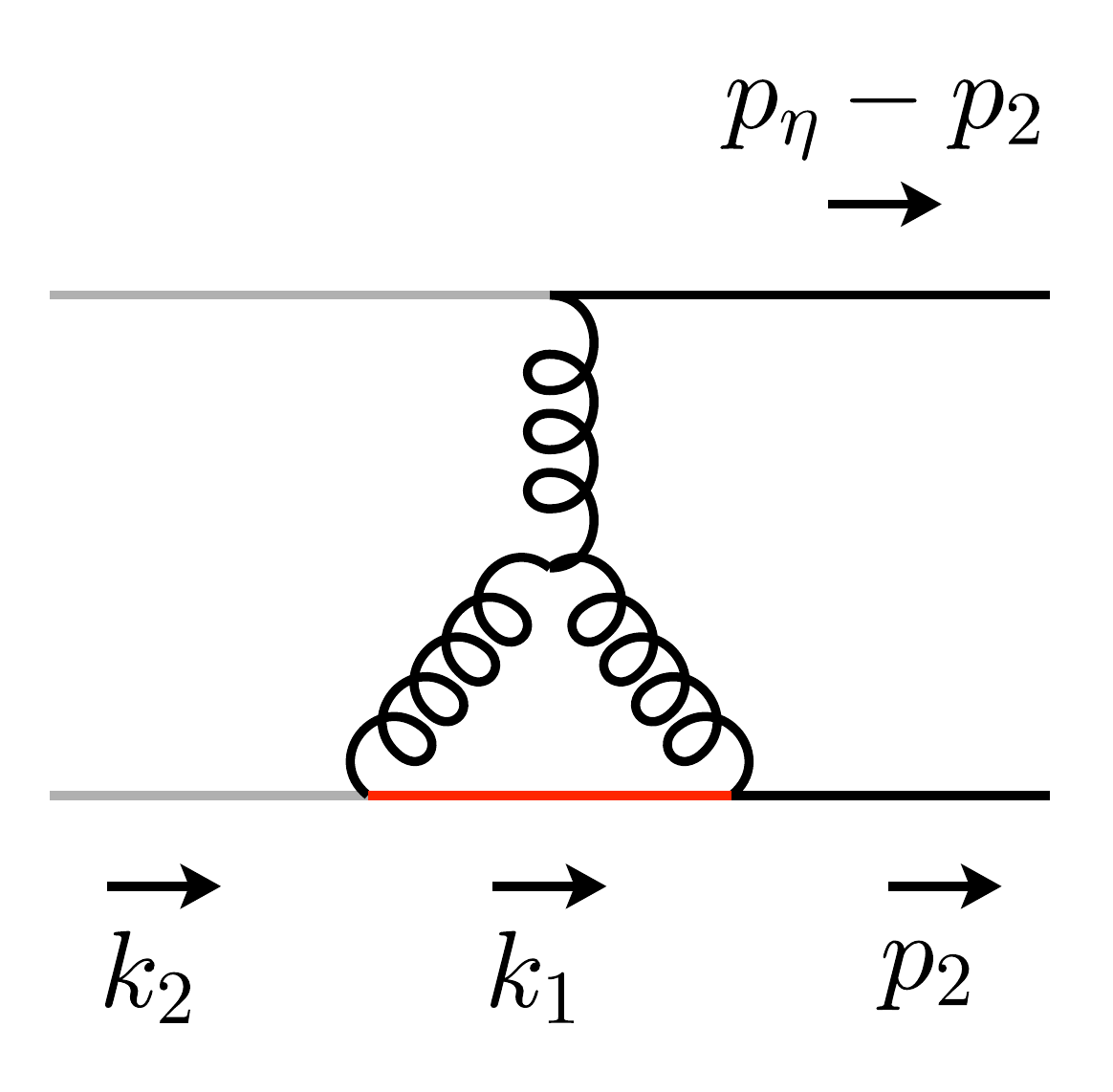}}} \;\approx\; \int \!\! \frac{d^4k_1}{(2\pi)^4} \; 
	\frac{\text{Numerator}}{[k_1^2 - m_2^2] \, (-k_{1+} p_{2-}) \, (-k_{2+} p_{2-}) \, (-k_{2+} k_{1-})} \,, 
\end{align}
which shows that the double-logarithmic enhancement is now associated with those terms in the numerator that are independent of $k_{1+}$ and $k_{1-}$. As the diagram itself arises at one-loop order, we denote the new set of coefficients by $\hat{c}_i^{(1)}$. We find $\hat{c}_1^{(1)}=\hat{c}_3^{(1)}=\hat{c}_5^{(1)}=\hat{c}_7^{(1)}=0$, along with
\begin{align}
	\label{eq:all-order:non-abelian:initial}	
	\hat{c}_2^{(1)} = -\frac14 \hat{c}_4^{(1)} = - \frac13 \hat{c}_6^{(1)} \simeq 
	\frac{\pi \alpha_s C_F f_{\eta}}{N_c \gamma \bar{u}_0} \;\frac{\alpha_s C_A}{4\pi} 
	\int_{k_{2-}}^{p_{2-}} \! \frac{dk_{1-}}{k_{1-}}
	\int_{m_2^2/k_{1-}}^{k_{2+}} \!\! \frac{dk_{1+}}{k_{1+}}\,.
\end{align}
For a generic insertion of a three-gluon vertex, it is sufficient to consider the diagram with a non-Abelian interaction right after the first rung of the ladder,  since its output \eqref{eq:all-order:ladder:initial} takes the most general form. As this corresponds to a two-loop diagram, the index of the new coefficients must change by two units from $c_i^{(0)}$ to $\hat{c}_i^{(2)}$ in this case. By looking at the denominator structure in the usual hierarchy \eqref{eq:strong-ordering} of the relevant momenta,
\begin{align}		
	& \vcenter{\hbox{\vspace{+0.1cm} \includegraphics[trim = 18 402 18 18,clip,height=3cm]{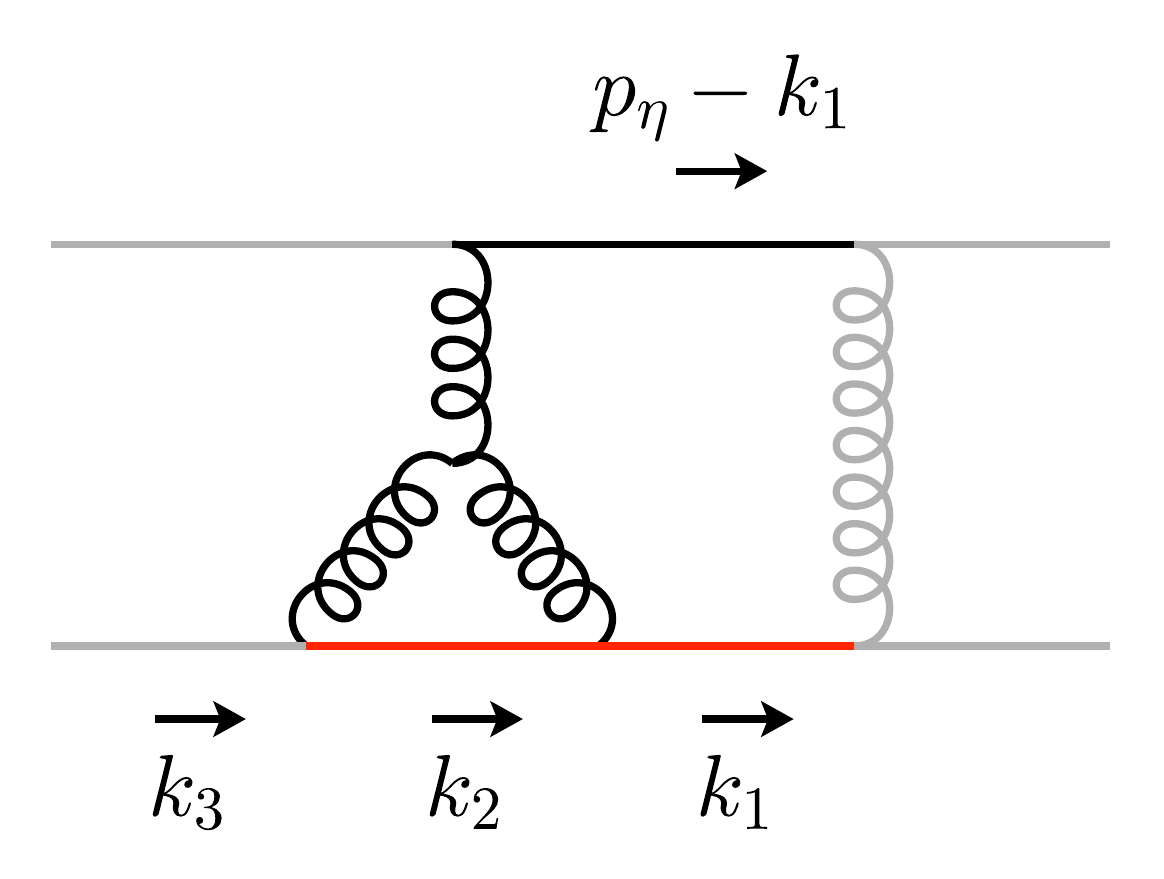}}}\\
	&\qquad \approx\; \int \!\! \frac{d^4k_1}{(2\pi)^4} \frac{d^4k_2}{(2\pi)^4}\; 
	\frac{\text{Numerator}}{[k_1^2 - m_2^2] \, [k_2^2 - m_2^2] \, (-k_{1+} p_{\eta-}) \, (-k_{2+} k_{1-}) \, (-k_{3+} k_{1-}) \, (-k_{3+} k_{2-}) \, k_{1+}} \,, \nonumber
\end{align}
one observes that the numerator must provide a factor of $k_{1+}k_{1-}$ for a leading-logarithmic sensitivity in this case. The explicit calculation yields
$\hat{c}_1^{(2)} = \hat{c}_3^{(2)} = \hat{c}_5^{(2)}  = \hat{c}_7^{(2)} =0$ and
\begin{align}
	\label{eq:all-order:non-abelian:coefficients}
	\left. \begin{array}{l}
		\hat{c}_2^{(2)} \\[0.5em]
		\hat{c}_4^{(2)} \\[0.5em]
		\hat{c}_6^{(2)} 
	\end{array} \right\} \, \simeq
	\, \Big(\frac{\alpha_s}{4\pi}\Big)^2 C_A C_F
	\int_{k_{3-}}^{p_{2-}} \! \frac{dk_{2-}}{k_{2-}} 
	\int_{k_{2-}}^{p_{2-}} \! \frac{dk_{1-}}{k_{1-}}
	\int_{m_2^2/k_{2-}}^{k_{3+}} \!\! \frac{dk_{2+}}{k_{2+}} 	
	\int_{m_2^2/k_{1-}}^{k_{2+}} \!\! \frac{dk_{1+}}{k_{1+}} \,
	\left\{ \begin{array}{l}
		-2\bar{u}_0 \, c_1^{(0)} + 2 c_3^{(0)}, \\[0.5em]
		8\bar{u}_0 \, c_1^{(0)} -8 c_3^{(0)}, \\[0.5em]
		6\bar{u}_0 \, c_1^{(0)} -6 c_3^{(0)}. 
	\end{array} \right.
\end{align}   
As the input to this two-loop integral was generic, the procedure can again be iterated to higher orders with coefficients $\hat{c}_i^{(j)}$ expressed through lower-order coefficients $c_i^{(j-2)}$ via \eqref{eq:all-order:non-abelian:coefficients}. Moreover, the pattern \eqref{eq:all-order:non-abelian:coefficients} also applies to cases with two non-Abelian insertions next to each other, i.e.~with hatted coefficients on the right-hand side of this equation. Similarly, the above equations \eqref{eq:all-order:ladder:coefficients} and \eqref{eq:all-order:final-projection} can also be used with a non-Abelian contribution as input and hatted coefficients on their right-hand sides. With this set of generalized iteration equations, one then easily verifies that this procedure correctly reproduces the contributions from the non-Abelian one-loop and two-loop diagrams in \eqref{eq:NLO:D2:result} and \eqref{eq:NNLO:D2pD3p:result}, respectively. \\

The structure of the iteration equations \eqref{eq:all-order:ladder:coefficients} and \eqref{eq:all-order:non-abelian:coefficients} has important implications. First of all, we recall that only the coefficients $c_i^{(n)}$ with $i=1,2,3$ contribute in the final projection \eqref{eq:all-order:final-projection}. As we have seen in \eqref{eq:all-order:ladder:coefficients}, these coefficients mix among themselves under an Abelian interaction, and the same is true in the non-Abelian case as well. We may therefore concentrate on this subset of coefficients from now on, which could have been anticipated from closer inspecting the basis elements in \eqref{eq:all-order:basis}, since a factor $\slashed{n}$ to the right necessarily leads to a power suppression, and the fifth basis element is suppressed by a factor of $1/\gamma$. Moreover, the above equation \eqref{eq:all-order:non-abelian:coefficients} reveals that the non-Abelian diagrams only take $c_{i}^{(j-2)}$ coefficients with $i=1,3$ as input, which generate a $\hat{c}_{2}^{(j)}$ term. Likewise, the right-most non-Abelian insertion only produces a $\hat{c}_{2}^{(1)}$ term as well, see \eqref{eq:all-order:non-abelian:initial}. This implies that two non-Abelian insertions next to each other cannot produce a leading-logarithmic contribution, since their respective input and output do not match. But by closer inspecting \eqref{eq:all-order:ladder:coefficients}, we see that the $c_{2}^{(j)}$ coefficients mix only into themselves under Abelian ladder exchanges, and we can therefore conclude that the relevant diagrams can have at most a \emph{single} non-Abelian insertion at any order in perturbation theory. This observation goes beyond the fixed-order analysis from the previous section, since a diagram with two non-Abelian insertions first arises at three-loop order. 

We may further simplify the iteration equations by noting that the $c_{1}^{(j)}$ coefficients only contribute at tree level with $j=0$. It is therefore instructive to remove these coefficients with a basis change,
\begin{align}
\label{eq:all-order:primed-basis}
	c_2^{\prime\,(j)} = c_2^{(j)} - c_1^{(j)}\,,
    \qquad\qquad
    c_3^{\prime\,(j)} = c_3^{(j)} - \bar{u}_0 \,c_1^{(j)}\,, 
\end{align} 
such that the new set of primed coefficients satisfy the same equations as given in \eqref{eq:all-order:ladder:coefficients} and \eqref{eq:all-order:non-abelian:coefficients} for the unprimed coefficients, but without the $c_{1}^{(j)}$ terms. At tree level, one then has 
\begin{align}
\label{eq:all-order:tree-level:primed}
c_2^{\prime\,(0)} = 0\,,
\qquad\qquad
c_3^{\prime\,(0)} =
\frac{\pi \alpha_s C_F f_{\eta}}{N_c \gamma \bar{u}_0} \;(1+\bar{u}_0)\,, 
\end{align}   
in this notation. Moreover, it is desirable to combine the Abelian and non-Abelian iterations into a unified set of equations, which is not a completely obvious task, since the Abelian iteration proceeds in one step, whereas the non-Abelian iteration relates two orders. We just noted, however, that the non-Abelian insertions can only occur once, and one may therefore rephrase the non-Abelian iteration step in a way that links $\hat{c}_2^{\prime\,(j)}$ to $c_3^{\prime\,(j-1)}$ via a $C_A$-type contribution, letting the usual Abelian iteration from $c_3^{\prime\,(j-1)}$ to $c_3^{\prime\,(j-2)}$ account for the remaining factors and integrals. This yields the following compact set of iteration equations
\begin{align}
\label{eq:all-order:primed:coefficients} 
	\left. \begin{array}{l}
		c_2^{\prime\,(j)} \\[0.5em]
		c_3^{\prime\,(j)} 
	\end{array} \right\} \simeq
	\frac{\alpha_s}{4\pi} 
	\int_{k_{(j+1)-}}^{p_{2-}} \! \frac{dk_{j-}}{k_{j-}}
	\int_{m_2^2/k_{j-}}^{k_{(j+1)+}} \! \frac{dk_{j+}}{k_{j+}} 
    \left\{ \begin{array}{l} \displaystyle
	2C_F \, c_2^{\prime\,(j-1)} +(C_A - 2C_F) c_3^{\prime\,(j-1)} 
    - \delta_{j1} C_A \,\frac{\bar{u}_0 \,c_3^{\prime\,(0)}}{1+\bar{u}_0}, \\[0.5em]
	2C_F \, c_3^{\prime\,(j-1)},
	\end{array} \right. 
\end{align} 
which supersede \eqref{eq:all-order:ladder:coefficients} and \eqref{eq:all-order:non-abelian:coefficients}. The last term in the first line requires some explanation, since our analysis was based on \eqref{eq:all-order:non-abelian:coefficients}, which is valid for $j\geq 2$. We are free, however, to bring the contribution from the right-most non-Abelian insertion in \eqref{eq:all-order:non-abelian:initial} into the same form by adding a suitable inhomogeneity to this equation that only contributes for $j=1$. The above equations are therefore valid for $j\geq 1$, with the corresponding initial values given in \eqref{eq:all-order:tree-level:primed}. After translating the final projection identity \eqref{eq:all-order:final-projection} into the new basis \eqref{eq:all-order:primed-basis}, this then determines the leading-logarithmic soft-quark contribution to all orders in perturbation theory. 

Instead of working through the iteration equations \eqref{eq:all-order:primed:coefficients} order-by-order, one may intro\-duce two auxiliary functions that capture the leading-logarithmic contribution to the soft-overlap form factor to all orders. This reformulation will also be beneficial for the discussion of soft-gluon corrections in the following section. Specifically, one notes that the iteration pattern for the $c_3^{\prime\,(j)}$ coefficients is familiar from muon-electron backward scattering~\cite{Gorshkov:1966qd,Bell:2022ott}. The recursively defined function\footnote{Note that the function $f_1$ was denoted $f_m$ in~\cite{Boer:2023tcs}, and similarly the function $f_2$ from \eqref{eq:recursion-relation:f:soft-quark} below was simply called $f$ in that reference.}
\begin{align} 
\label{eq:recursion-relation:fm:soft-quark}	
  f_1(q_+,q_-) &= 1 + \frac{\alpha_s}{4\pi} \int\limits_{q_-}^{p_{2-}} \frac{dk_-}{k_-} \! \int\limits_{m_{2}^2/k_-}^{q_+} \!\frac{dk_+}{k_+} \,  \Big[2 C_F  \, f_1(k_+,k_-) \Big] 
\end{align}
generates the series of $c_3^{\prime\,(j)}$ coefficients when evaluated for on-shell kinematics $q_+=\ell_+=m_2$ and $q_-=\ell_-=m_2$ according to 
\begin{align} 
 \sum_{j=0}^\infty \,c_3^{\prime\,(j)} =
\frac{\pi \alpha_s C_F f_{\eta}}{N_c \gamma \bar{u}_0} \; (1+\bar{u}_0)\,
f_1(m_2,m_2)\,. 
\end{align}
Likewise, we define a second function recursively by
\begin{align} 
\label{eq:recursion-relation:f:soft-quark}	
  f_2(q_+,q_-) &= 1 + \frac{\alpha_s}{4\pi} \int\limits_{q_-}^{p_{2-}} \frac{dk_-}{k_-} \! \int\limits_{m_{2}^2/k_-}^{q_+} \!\frac{dk_+}{k_+} \,  \bigg[2 C_F  \, f_2(k_+,k_-) 
  + \Big( C_F - \frac{C_A}{2} \Big) f_1 (k_+,k_-)  + \frac{C_A}{2} \bigg] \, , 
\end{align}
which was designed to generate the series that enters in the final projection \eqref{eq:all-order:final-projection} in the new basis via
\begin{align} 
 \sum_{j=0}^\infty \, \big( 2 c_3^{\prime\,(j)} - c_2^{\prime\,(j)} \big) =
\frac{\pi \alpha_s C_F f_{\eta}}{N_c \gamma \bar{u}_0} \; \bigg[ 2 (1+\bar{u}_0)\,
f_2(m_2,m_2) + \frac{C_A}{2C_F} \, \big(1- f_1(m_2,m_2)\big) \bigg]\,. 
\end{align}
Expressed in terms of these auxiliary functions, the pure soft-quark contribution to the soft-overlap form factor can then be written to all orders in perturbation theory in the form
\begin{equation}
	\label{eq:all-order:soft-quark:result}	
F_{\text{soft quarks}}(\gamma)  \simeq \,
	\xi_0 \left(2 \frac{1+\bar{u}_0}{\bar{u}_0^3}\,
f_2(m_2,m_2) + \frac{C_A}{2C_F \bar{u}_0^3} \, \big(1- f_1(m_2,m_2)\big)  -\frac{1}{\bar{u}_0^2}\right). 
\end{equation}
The structure of this equation can be understood as follows. First of all, we note that there are different contributions in the quark-mass ratio $\bar{u}_0$. In particular, all Abelian ladder diagrams contribute via the first term in \eqref{eq:all-order:soft-quark:result} and are therefore proportional to $(1+\bar{u}_0)/\bar{u}_0^3$, except for the tree-level diagram itself, which receives a correction from the non-vanishing $c_1^{(0)}$ coefficient that generates the last term in \eqref{eq:all-order:soft-quark:result}. This is in line with our findings in \eqref{eq:xiLO}, \eqref{eq:NLO:D1:result} and \eqref{eq:NNLO:D1p:result}. Moreover, there are two different classes of non-Abelian contributions signalled by the presence of $C_A$ terms in \eqref{eq:all-order:soft-quark:result} and \eqref{eq:recursion-relation:f:soft-quark}. The first class in \eqref{eq:all-order:soft-quark:result} captures the contribution from the diagrams with right-most insertions of a three-gluon vertex, and from \eqref{eq:NLO:D2:result} and the first equation in \eqref{eq:NNLO:D2pD3p:result} we indeed see that these are proportional to $1/\bar{u}_0^3$. The $C_A$ terms in \eqref{eq:recursion-relation:f:soft-quark} then generate all remaining non-Abelian contributions (with a single insertion of a three-gluon vertex), and the second relation in \eqref{eq:NNLO:D2pD3p:result} confirms that these come again with a factor $(1+\bar{u}_0)/\bar{u}_0^3$. As these diagrams only start at $\mathcal{O}(\alpha_s^2)$, the $C_A$ term in \eqref{eq:recursion-relation:f:soft-quark} has to cancel out at $\mathcal{O}(\alpha_s)$, which is precisely achieved by the last term in the parenthesis in \eqref{eq:recursion-relation:f:soft-quark}.

The implicit integral equations~\eqref{eq:recursion-relation:fm:soft-quark} and~\eqref{eq:recursion-relation:f:soft-quark} can be solved in closed form in terms of modified Bessel functions~\cite{Gorshkov:1966qd,Bell:2022ott,Boer:2023tcs}. Defining $z = \frac{\alpha_s C_F}{2\pi} L^2$, we find for the two functions that enter the soft-overlap form factor in~\eqref{eq:all-order:soft-quark:result},
\begin{align}
\label{eq:puresoftquarks}
    f_1(m_2,m_2) \, &\stackrel{\text{soft quarks}}{=} \, \frac{I_1(2\sqrt{z})}{\sqrt{z}} \,, \nonumber \\
    f_2(m_2,m_2) \, &\stackrel{\text{soft quarks}}{=} \, \frac{C_F+C_A}{2C_F} \, \frac{I_1(2\sqrt{z})}{\sqrt{z}} + \frac{2C_F - C_A}{4C_F} \, I_0(2\sqrt{z})  - \frac{C_A}{4 C_F} \,,
\end{align}
and we remind the reader that these expressions only resum the pure soft-quark corrections.

\subsection{Interplay with soft gluons}
\label{sec:all-order:interplay}

Having understood the soft-quark dynamics to all orders, we will now address the second source of leading-logarithmic corrections from soft-gluon exchanges. To this end, we first note that the spin-independent (eikonal) soft-gluon couplings cannot change the structure of the soft-quark mixing encoded in the coupled integral equations \eqref{eq:recursion-relation:fm:soft-quark} and \eqref{eq:recursion-relation:f:soft-quark}. Moreover, soft-gluon corrections are known to exponentiate, and they are proportional to the Casimir of the emitting particle. Still, we saw in Sec.~\ref{subsec:NNLO} that the interplay of soft-quark and soft-gluon corrections is non-trivial, since the phase space of the soft-gluon emissions can depend on the soft-quark configuration. This modifies the integrands of the soft-quark contributions, and we need to understand how this effect can be accounted for to all orders in our setup. 

To approach this question, we recall how the Sudakov factor is generated at NNLO, cf.~in particular the discussion around the diagrams in Fig.~\ref{fig:NNLO:interplay:diagrams} in hybrid gauge. Whereas the first diagram yields an overall factor $-\frac{\alpha_s C_F}{4\pi} L^2$, we found that the remaining diagrams conspire to $-s(\ell_+/k_+,p_{\eta -}/k_-)$, where we introduced the Sudakov exponent 
\begin{equation}
    s(r_+,r_-) = \frac{\alpha_s C_F}{2\pi} \, \ln r_+ \, \ln r_- \,. 
\end{equation}
The non-trivial task then consists in deriving the phase-space boundaries of the soft-gluon emissions -- which translate into the arguments of the Sudakov exponent -- for a generic soft-quark diagram. As is evident from Fig.~\ref{fig:NNLO:interplay:diagrams}, these phase-space boundaries can be most easily read off from the non-Abelian diagram $D_8'$ at NNLO, although the diagram itself comes with the wrong color factor. In other words, it is possible to extract the arguments of the Sudakov exponent in the large $N_c$ limit, which significantly simplifies the diagrammatic analysis.

\begin{figure}[t]
    \centering
    \includegraphics[trim = 18 617 18 18,clip,height=0.15 \textheight]{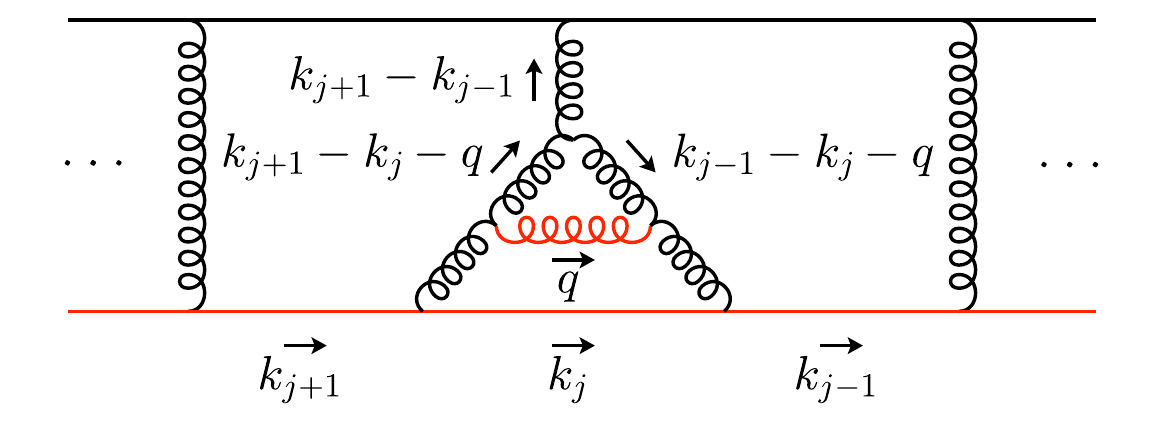}
    \vspace{-0.2cm}
    \caption{Planar soft-gluon attachment to a generic non-Abelian loop in the soft-quark ladder.}
\label{fig:interplay:non-Abelian}
\end{figure}

Let us now consider a generic insertion of a three-gluon vertex in the soft-quark ladder, as illustrated in Fig.~\ref{fig:interplay:non-Abelian}. In order to understand how the soft-quark loop with momentum $k_j$ is altered by soft-gluon effects, we consider the corresponding planar soft-gluon attachment with a momentum assignment as shown in the figure. Eikonalizing the relevant propagators then gives the constraints
\begin{align}
    (k_{j+1}-k_j-q)^2 \,\approx\, - k_{(j+1) +} q_- \qquad
    \Rightarrow \qquad q_- > k_{j-} \,,
    \qquad q_+ < k_{(j+1)+} \,, \cr 
    (k_{j-1}-k_j-q)^2 \,\approx\, - k_{(j-1) -} q_+ \qquad
    \Rightarrow \qquad q_+ > k_{j+} \,,
    \qquad q_- < k_{(j-1)-} \,, 
\end{align}
which leads to the integral
\begin{align}
\label{eq:all-order:interplay:non-Abelian:Sudakov}
	- \frac{\alpha_s C_F}{2\pi} 
    \int_{k_{j+}}^{k_{(j+1)+}} \! \frac{dq_+}{q_+}  
    \int_{k_{j-}}^{k_{(j-1)-}} \! \frac{dq_-}{q_-} 
	\,=  \, - \frac{\alpha_s C_F}{2\pi}  \,
    \ln \bigg(\frac{k_{(j+1)+}}{k_{j+}}\bigg)
    \ln \bigg(\frac{k_{(j-1)-}}{k_{j-}}\bigg)\,. 
\end{align}
This is consistent with the factor $-s(\ell_+/k_+,p_{\eta -}/k_-)$ we found at NNLO, since the adjacent momenta in the two-loop diagram can be identified with $k_{(j+1)+}=\ell_+$ and $k_{(j-1)-}=p_{2-}$ in this case, and the distinction between $p_{2-}$ and $p_{\eta-}$ is irrelevant in the double-logarithmic approximation. Interestingly, we thus see that the arguments of the Sudakov exponent are not fixed by the external kinematics, but by the neighbouring loop momenta.

\begin{figure}[t]
    \centering
    \includegraphics[trim = 18 565 18 18,clip,height=0.18 \textheight]{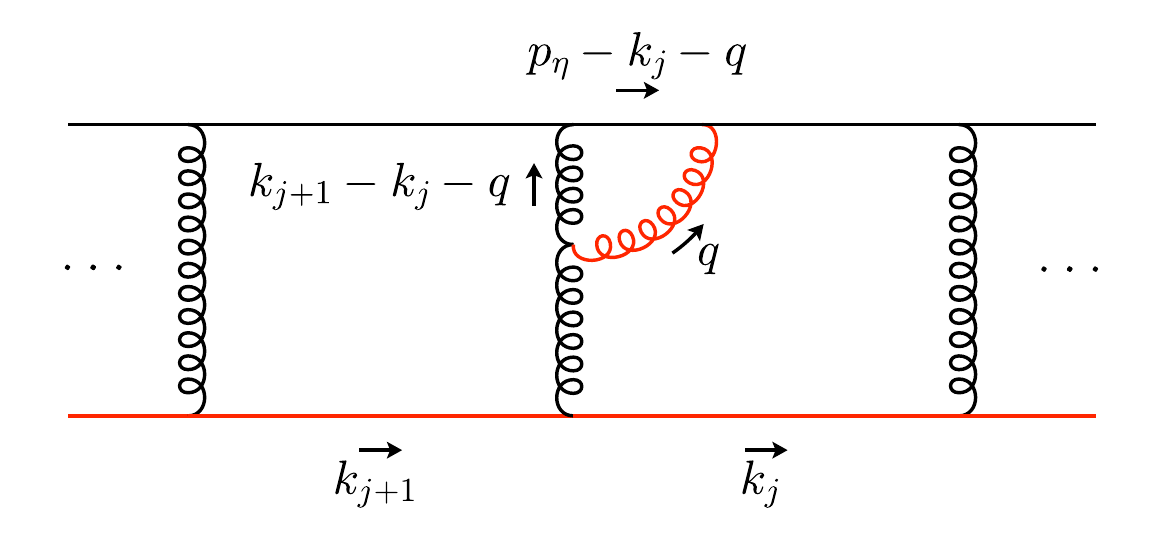}
    \vspace{-0.2cm}
    \caption{Planar soft-gluon attachment to a generic rung in the soft-quark ladder.}
\label{fig:interplay:ladder}
\end{figure}

We can perform a similar analysis for the second building block of the soft-quark dia\-grams, i.e.~the Abelian ladder exchanges. Although we did not work through the detailed diagrammatic analysis in Sec.~\ref{subsec:NNLO}, it is also true in this case that the corresponding Sudakov exponent can be extracted most efficiently in the large $N_c$ limit. The relevant planar soft-gluon attachment is shown in Fig.~\ref{fig:interplay:ladder} for a generic ladder diagram. Arguing as before in the eikonal approximation, one now has
\begin{align}
    (k_{j+1}-k_j-q)^2 \,\approx\, - k_{(j+1) +} q_- &\qquad
    \Rightarrow \qquad q_- > k_{j-} \,,
    \qquad q_+ < k_{(j+1)+} \,,  \cr
    (p_\eta-k_j-q)^2 \,\approx\, - p_{\eta-} q_+ &\qquad
    \Rightarrow \qquad q_+ > k_{j+} \,,
    \qquad q_- < p_{\eta-} \,, 
\end{align}
which translates into
\begin{align}
\label{eq:all-order:interplay:ladder:Sudakov}
	- \frac{\alpha_s C_F}{2\pi} 
    \int_{k_{j+}}^{k_{(j+1)+}} \! \frac{dq_+}{q_+}  
    \int_{k_{j-}}^{p_{\eta-}} \! \frac{dq_-}{q_-} 
	\,=  \, - \frac{\alpha_s C_F}{2\pi}  \,
    \ln \bigg(\frac{k_{(j+1)+}}{k_{j+}}\bigg)
    \ln \bigg(\frac{p_{\eta-}}{k_{j-}}\bigg)\,. 
\end{align}
Taking into account that $k_{(j+1)+}=\ell_+$ for the one-loop box diagram $D_1$ from Fig.~\ref{fig:NLO:diagrams}, this again reduces to $-s(\ell_+/k_+,p_{\eta -}/k_-)$  at NNLO. The generic case shows, however, that the two relevant building blocks of the soft-quark diagrams come with \emph{different} Sudakov exponents given by \eqref{eq:all-order:interplay:non-Abelian:Sudakov} and \eqref{eq:all-order:interplay:ladder:Sudakov}, and it is a priori not clear how this information can be embedded into the integral equations \eqref{eq:recursion-relation:fm:soft-quark} and \eqref{eq:recursion-relation:f:soft-quark}, where Abelian and non-Abelian corrections are entangled in a non-trivial way. 

In order to clarify this point, we first note that for right-most non-Abelian insertions, which are described by the $C_A$-terms in \eqref{eq:all-order:soft-quark:result}, one again has $k_{(j-1)-}=p_{2-}$ and the two Sudakov factors from \eqref{eq:all-order:interplay:non-Abelian:Sudakov} and \eqref{eq:all-order:interplay:ladder:Sudakov} thus coincide in the double-logarithmic approximation. For a generic non-Abelian insertion, on the other hand, the key observation is that the loop to the right of the non-Abelian insertion is not generic. To be specific, consider once again the momentum assignment of the diagram shown in Fig.~\ref{fig:interplay:non-Abelian}. The relevant gluon line that is responsible for the Sudakov factor associated with the loop momentum $k_{j-1}$ then carries a momentum $(k_{j+1}-k_{j-1})$, which is different from the combination $(k_{j}-k_{j-1})$ that appears in generic ladder diagrams. Spelling out the two Sudakov exponents of the non-Abelian insertion and the special ladder diagram next to it then yields
\begin{align}
& - \frac{\alpha_s C_F}{2\pi}  \bigg\{
    \underbrace{\ln \bigg(\frac{k_{(j+1)+}}{k_{j+}}\bigg)
    \ln \bigg(\frac{k_{(j-1)-}}{k_{j-}}\bigg)}_{\text{non-Abelian insertion}}
    +
    \underbrace{\ln \bigg(\frac{k_{(j+1)+}}{k_{(j-1)+}}\bigg)
    \ln \bigg(\frac{p_{\eta-}}{k_{(j-1)-}}\bigg)}_{\text{special ladder insertion}}
    \bigg\}\nonumber\\[0.8em]
& \qquad = - \frac{\alpha_s C_F}{2\pi}  \bigg\{
    \ln \bigg(\frac{k_{(j+1)+}}{k_{j+}}\bigg)
    \ln \bigg(\frac{p_{\eta-}}{k_{j-}}\bigg)
    +
    \ln \bigg(\frac{k_{j+}}{k_{(j-1)+}}\bigg)
    \ln \bigg(\frac{p_{\eta-}}{k_{(j-1)-}}\bigg)
    \bigg\}\,. 
\end{align}
In the combination of the two insertions, we thus obtain a Sudakov factor that is precisely of the form \eqref{eq:all-order:interplay:ladder:Sudakov} for each of the loop momenta. This implies that the all-order soft-gluon effect can be implemented straight-forwardly into the integral equations \eqref{eq:recursion-relation:fm:soft-quark} and \eqref{eq:recursion-relation:f:soft-quark}, since each loop in the soft-quark chain comes with a factor $\exp{-s(k_{(j+1)+}/k_{j+},p_{\eta -}/k_{j-})}$ for $j=1,\ldots,n$, if we identify $k_{n+1} = \ell$. Translated into the notation we introduced in the previous section, where $k$ and $q$ denote two adjacent momenta of the spectator antiquark, this then leads to the final form of the coupled integral equations,
\begin{align} 
  f_1(q_+,q_-) &= 1 + \frac{\alpha_s}{4\pi} \int\limits_{q_-}^{p_{2-}} \frac{dk_-}{k_-} \! \int\limits_{m_{2}^2/k_-}^{q_+} \!\frac{dk_+}{k_+} \;  
  e ^{-s(q_+/k_+,p_{\eta -}/k_-)} \,
  \Big[2 C_F  \, f_1(k_+,k_-) \Big] \,,
    \cr
f_2(q_+,q_-) &= 1 + \frac{\alpha_s}{4\pi} \int\limits_{q_-}^{p_{2-}} \frac{dk_-}{k_-} \! \int\limits_{m_{2}^2/k_-}^{q_+} \!\frac{dk_+}{k_+} \;  
  e ^{-s(q_+/k_+,p_{\eta -}/k_-)} 
  \cr
  & \hspace{20mm} \times 
  \bigg[2 C_F  \, f_2(k_+,k_-) 
  + \Big( C_F - \frac{C_A}{2} \Big) f_1 (k_+,k_-)  + \frac{C_A}{2} \bigg] \, . 
\label{eq:finalintegralequations}
\end{align}
In addition to this, the overall Sudakov factor modifies the relation \eqref{eq:all-order:soft-quark:result} to
\begin{equation}
\label{eq:Ffinaldecomp}
F(\gamma)  \simeq \,
	\xi_0 \,\exp\bigg\{ -\frac{\alpha_s C_F}{4\pi} L^2\bigg\} 
    \left(2 \frac{1+\bar{u}_0}{\bar{u}_0^3}\,
f_2(m_2,m_2) + \frac{C_A}{2C_F \bar{u}_0^3} \, \big(1- f_1(m_2,m_2)\big)  -\frac{1}{\bar{u}_0^2}\right). 
\end{equation}
The two equations \eqref{eq:finalintegralequations} and \eqref{eq:Ffinaldecomp} are the main result of our analysis. They reflect an intricate interplay of soft-quark and soft-gluon corrections, and determine the full tower of leading-logarithmic corrections to the soft-overlap form factor to all orders in perturbation theory. It then becomes a straight-forward task to verify that they reproduce the known fixed-order results from \eqref{eq:xiLO}, \eqref{eq:NLO:DL} and \eqref{eq:NNLO:full:result} up to NNLO, and they can in principle be iterated to even higher orders. We will study the structure of the integral equations and their solution in more detail in the following section.

\section{Analysis of the integral equations}
\label{sec:integraleqs}

In the previous section we established a novel type of implicit integral relations~\eqref{eq:finalintegralequations} for the auxiliary form factors $f_{1,2}(q_+,q_-)$, which encode a non-trivial interplay of the soft-quark and soft-gluon dynamics in the soft-overlap form factor. In the following, we discuss how to construct the solutions to these equations in an iterative way, leading to the summation of the associated double-logarithmic QCD corrections to any desired order in perturbation theory. In addition, we derive the asymptotic behaviour of the form factors in the limit of infinitely large recoil.

To begin with, we remind the reader that the solutions to the integral equations need to be evaluated for the on-shell kinematics of the process, and we thus find it useful to define on-shell form factors appearing in~\eqref{eq:Ffinaldecomp} as
\begin{align}
    G_1(z) \equiv f_1(m_2,m_2) \,, \qquad G_2(z) \equiv f_2(m_2,m_2) \,,
    \label{eq:ftoG}
\end{align}
with $ z = \frac{\alpha_s C_F}{2\pi} L^2$. 
For the following analysis, it also turns out to be convenient to introduce logarithmic variables, 
defined in terms of the boost-invariant ratios $q_+ p_-/m_2^2$ and $q_-/p_-$,
\begin{equation}
\rho = \sqrt{\frac{\alpha_s C_F}{2\pi}} \, \ln \left(\frac{q_+ p_-}{m_2^2}\right) \,, \qquad 
 \eta = \sqrt{\frac{\alpha_s C_F}{2\pi}} \,\ln \left(\frac{p_-}{q_-}\right) \,,
\end{equation}
and we denote
\begin{equation}
 f_1(q_+,q_-) = g_1(\rho,\eta) \,, \qquad f_2(q_+,q_-) = g_2(\rho,\eta) \,.
 \end{equation}
Notice that from here on, we do not distinguish between the logarithms of the light-cone momenta $p_{\eta -}$ and $p_{2-}$ of the energetic meson and quark anymore, and instead denote $p_- \approx p_{\eta_-} \approx p_{2-}$, which is correct up to single-logarithmic effects.
The integral equations then take the form
\begin{align} 
\label{eq:inteqslogvar}
 g_1(\rho,\eta) &= 1 + \int_0^\eta \! d\eta' \int_{\eta'}^{\rho} \! d\rho' \, e^{-\eta' \, (\rho-\rho')} \,  g_1(\rho',\eta') \,, \nonumber \\[2ex]
 g_2(\rho,\eta) &= 1 + \int_0^\eta \! d\eta' \int_{\eta'}^{\rho} \! d\rho' \, e^{- \eta' \, (\rho-\rho')} 
 \, \left[ \left( g_2(\rho',\eta') + \frac12 \right) -\frac{1}{4 N_c C_F} \big( g_1(\rho',\eta') - 1 \big) \right] \,,
\end{align}
where we have used that $C_A = 2C_F + 1/N_c$.
Notice that the non-trivial mixing between
the two functions $g_1(\rho,\eta)$ and $g_2(\rho,\eta)$ is suppressed by $1/(4N_c C_F) \sim 1/N_c^2$ in the large $N_c$ limit.

Before we analyze these relations further in the next subsection, it is instructive to compare them with two related results discussed recently in the literature: 
\begin{itemize}
    \item[(i)] Using an analogous notation as above, the leading double-logarithmic corrections to the amplitude of bottom-induced Higgs decays to photons reads~\cite{Kotsky:1997rq,Liu:2018czl,Liu:2019oav}
    \begin{align}
        {\cal A}^{(b)}(h \to \gamma\gamma) 
        &\simeq \frac{2\pi  {\cal A}_0^{(b)}}{\alpha_s C_F} \, \int_0^\eta d\eta' \, \int_{\eta'}^\rho d\rho' \, e^{- \eta' \, (\rho-\rho') } \,, \quad \mbox{$\rho = \eta = \sqrt{\frac{\alpha_s C_F}{2\pi} \, \ln^2 \frac{m_H^2}{m_b^2}}$} \,.
    \end{align}
In this case, the soft-quark logarithms from the on-shell bottom-quark propagator do not iterate (more precisely, iterated soft-bottom propagators are suppressed by powers of the bottom mass), whereas double logarithms associated with soft gluons from the hard $h\to b\bar b$ vertex exponentiate in the usual way. Therefore, the integration can be performed explicitly, and the result can be expressed in terms of a generalized hypergeometric function
\begin{align} 
  \int_0^\rho d\eta' \, \int_{\eta'}^\rho d\rho' \, e^{- \eta' \, (\rho-\rho') } 
  &= \frac{\rho^2}{2} \ {}_2F_2\big(1,1;\frac32,2;- \frac{\rho^2}{4}\big) \,.
\end{align}  

\item[(ii)] A complementary situation arises in muon-electron scattering in the high-energy limit and exact backward kinematics~\cite{Bell:2022ott}. Here, the scattering amplitude can be described by a single form factor $f_1$ in the leading double-logarithmic approximation, which fulfills an integral relation of the form
\begin{align}
    f_1(\rho,\eta) &=1+ \int_0^\eta d\eta' \int_{\eta'}^\rho d\rho' \, f_1(\rho',\eta') \,. 
\end{align}
In this case, there is no \emph{external} momentum transfer, and therefore the Sudakov factor in the integrand is absent. On the other hand, the iteration of soft-lepton propagators in (crossed) ladder diagrams does not lead to a power suppression, and hence \emph{all} double-logarithmic corrections arise from configurations with rapidity-ordered and on-shell soft-lepton propagators. This integral relation can again be solved in a closed form, and evaluating the form factor for on-shell kinematics $\rho = \eta$ results in a modified Bessel function,
\begin{align}
\label{eq:BesselI1}
    f_1(\rho,\rho) &= \frac{I_1(2\rho)}{\rho} \,, \qquad 
    \mbox{with $\rho= \sqrt{\frac{\alpha}{2\pi} \, \ln^2 \frac{s}{m^2}}$} \,,
\end{align}
with $m$ the lepton mass and $s = 4E^2$ given by the center-of-mass energy.
\end{itemize}
In our case, we face the more general situation where both sources of large double logarithms, 
from soft-gluon and \textit{iterated} soft-quark configurations, enter in an intertwined way.

\subsection{One-fold integral equations for on-shell form factors}
\label{subsec:onefoldinteqs}

By taking derivatives with respect to the variables $(\rho,\eta)$, the integral equations~\eqref{eq:inteqslogvar} can be cast into a coupled system of hyperbolic partial differential equations
\begin{align}
    \left( \partial_\rho \partial_\eta + \eta \partial_\eta - 1 \right) g_1(\rho,\eta) &= 0 \,, \nonumber \\
    \left( \partial_\rho \partial_\eta + \eta \partial_\eta - 1 \right) \left( g_2(\rho,\eta) +\frac12 \right) &=
     - \frac{1}{ 4N_c C_F} \left( g_1(\rho,\eta) - 1 \right) \,,
    \label{eq:pdes}
\end{align}
with boundary conditions
\begin{align}
    g_1(\rho,\eta=0) &= 1 \,, & \partial_\eta g_1(\rho,\eta)\big\vert_{\rho = \eta} = 0 \,, \nonumber \\
    g_2(\rho,\eta=0) &= 1 \,, & \partial_\eta g_2(\rho,\eta)\big\vert_{\rho = \eta} = 0 \,.
\end{align}
Despite the apparently simple structure, we were not able to solve this set of partial differential equations in closed analytic form with standard mathematical methods, which can be traced back to the particular form of the boundary conditions.

Nevertheless, taking yet another derivative in the variable $\eta$ leads to rather simple differential equations for the second derivatives of the two auxiliary form factors in $\eta$,
\begin{align}
    \left( \partial_\rho + \eta \right) \partial_\eta^2 g_1(\rho,\eta) &= 0 \,, \nonumber \\
    \left( \partial_\rho + \eta  \right) \partial_\eta^2 g_2(\rho,\eta) &= - \frac{1}{ 4N_c C_F} \, \partial_\eta g_1(\rho,\eta) \,.
\end{align}
Let us first consider the homogeneous equation in the first row, which can simply be integrated. Taking into account the boundary conditions yields
\begin{align}
\label{eq:offandonshellgm}
  g_1(\rho,\eta) &= 1 +  
  \int_0^\eta d\eta' \, \int_{\eta'}^\rho d\eta'' \,
    e^{-  \eta'' \, (\rho-\eta'')} \, C_1(\eta'') 
    \nonumber \\
    &= 1 + \left(\int_0^\eta d\eta' \, \eta' + \int_\eta^\rho d\eta' \, \eta \right)
    e^{-\eta' \, (\rho-\eta')} \, C_1(\eta') \,,
\end{align} 
where the function $C_1$ has to be determined from the original equation. Indeed, from \eqref{eq:pdes} we obtain
\begin{align}
     g_1(\rho,\eta) &= \left( \partial_\rho + \eta  \right) \partial_\eta g_1(\rho,\eta) 
    \cr 
    &=  C_1(\rho) +  \int_\eta^\rho d\eta' \, (\eta-\eta') \, e^{-  \eta' \, (\rho-\eta')} \, C_1(\eta') \,.
\end{align}
From the latter equation, we immediately see that
\begin{align}
\label{eq:CmandGm}
    g_1(\rho,\rho) &= C_1(\rho) = G_1( \rho^2) \,, 
\end{align}
i.e.\ the function $C_1(\rho)$ determines the on-shell form factor $G_1(z) = f_1(m_2,m_2)$ entering~\eqref{eq:Ffinaldecomp}.  Furthermore, equating the two expressions for $g_1(\rho,\eta)$, gives a one-fold integral equation for the function $C_1(\rho)$,
\begin{align}
\label{eq:onefoldinteqs}
    C_1(\rho) &=
    1  + \int_0^\rho d\rho' \, \rho' \, e^{-  \rho' \, (\rho-\rho')} \, C_1(\rho') \,. 
\end{align}
Integrating over the interval $[0,\rho]$ the equation can be brought into the alternative form
\begin{align}
\label{eq:onefoldinteqs_integrated}
    \rho &= \int_0^\rho d\rho' \, e^{-\rho' \,(\rho-\rho')} \, C_1(\rho') \,,
\end{align}
which turns out to be more useful to discuss the asymptotic behaviour for $\rho \to \infty$, see below.
It is to be stressed that -- in contrast to the integral relation for the auxiliary form factor $g_1(\rho,\eta)$  -- the integral equation for $C_1(\rho)$ \emph{cannot} be translated to a simple differential equation, because each derivative in $\rho$ will generate an additional factor of $\rho'$ in the integrand on the right-hand side. Solving the equation iteratively order-by-order yields\footnote{The generating function for this series can be obtained by considering the Laplace transform of the function $h_1(x) \equiv e^{-x^2/2} \, C_1(ix)$. With (\ref{eq:onefoldinteqs}) one then obtains
$$
 {\cal L}[h_1](s) =  \frac{\sqrt{2}}{\sqrt{\pi}} \, \frac{e^{-s^2/2}}{{\rm erfc}(s/\sqrt2)} -s = 
 \frac{1}{s}-\frac{2}{s^3}+\frac{10}{s^5} - \frac{74}{s^7}+ \frac{706}{s^9} - \frac{8162}{s^{11}} + {\cal O}(s^{-13}) \,,
$$ with $\text{erfc}(s) = 1 - \text{erf}(s)$ being the complement of the error function.
The series expansion for $C_1(\rho)$ can then be obtained by the inverse Laplace transform and analytic continuation. 
} 
\begin{align}
  C_1(\sqrt z) = G_1(z) = \sum_{n=0}^\infty a_n \, (-z)^n 
    = 1 + \frac{z}{2} + \frac{z^2}{4!} - \frac{z^3}{6!} - \frac{z^4}{8!} + \frac{17 z^5}{10!} + \mathcal{O}(z^6)\,,
    \label{eq:Gmsmallz}
\end{align}
where the expansion coefficients obey the recursive relation\footnote{We remark in passing that the problem of finding a closed form for the expansion coefficients $a_n$ is related to the mathematical problem of finding the inverse of the binomial transform $g_n = \sum_{k=0}^n \binom{n+k}{k} \, f_k $, see \cite{Gould:1994}.}
\begin{align}
\label{eq:Gmcoefficientsrecursion}
    a_n &= -\sum_{k=0}^{n-1} \,\frac{(n+k)!}{(2n)!} \; a_k\,,
\end{align}
with $a_0=1$.

A similar analysis can be performed for the auxiliary form factor $g_2(\eta,\rho)$. It is convenient to disentangle the two color structures by defining two separate functions via
\begin{equation}
 g_2(\rho,\eta)\equiv g_{2a}(\rho,\eta) + \frac{1}{4N_c C_F} \, g_{2b}(\rho,\eta) \,,
\end{equation}
which satisfy
\begin{align}
     \left( \partial_\rho \partial_\eta + \eta \partial_\eta - 1 \right) \left( g_{2a}(\rho,\eta) +\frac12 \right) &=
     0 \,, \nonumber \\[0.2em]
      \left( \partial_\rho \partial_\eta + \eta \partial_\eta - 1 \right) g_{2b} (\rho,\eta)  &=
     -  \left( g_1(\rho,\eta) - 1 \right)\,, 
\end{align}
with $g_{2a}(\rho,0)=1$ and $g_{2b}(\rho,0)=0$. The form factor $g_{2a}(\rho,\eta)$ can be related to its on-shell limit via similar relations as~\eqref{eq:offandonshellgm} and~\eqref{eq:CmandGm}, with
\begin{align}
\label{eq:G1GM}
    g_{2a}(\rho,\rho) \equiv G_{2a}(\rho^2) &= \frac32 \, G_1(\rho^2)-\frac12 \,.
\end{align}
Performing the $\eta$-derivative of the second equation now yields
\begin{align}
 \left( \partial_\rho + \eta  \right) \partial_\eta^2 g_{2b}(\rho,\eta)= - \partial_\eta g_1(\rho,\eta) 
\end{align}
which is solved by
\begin{align}
\label{eq:g2ansatz}
  g_{2b}(\rho,\eta) &=  
  \int_0^\eta d\eta' \, \int_{\eta'}^\rho d\eta'' 
  e^{-  \eta'' \, (\rho-\eta'')} 
  \,
     C_{2b}(\rho,\eta'') 
    \,,
\end{align} 
where the $\rho$-dependence of the function $C_{2b}(\rho,\eta)$ is determined by the inhomogeneous term,
\begin{align}
    C_{2b}(\rho,\eta) & \equiv C_{2b}(\eta,\eta)   + \int_{\eta}^\rho d\rho' \, e^{-\eta(\eta-\rho')} 
     \, \partial_{\eta} g_1(\rho',\eta) 
     \cr &= C_{2b}(\eta,\eta)  +  \int_\eta^\rho d\eta' \int_{\eta'}^\rho d\rho' \, 
     e^{-\eta(\eta-\rho')} \, e^{-\eta'(\rho'-\eta')} \, C_1(\eta') \,. 
\end{align}
From the original differential equations we now obtain
\begin{align}
    g_{2b}(\rho,\eta) &= \left(\partial_\rho + \eta \right) \partial_\eta g_{2b}(\rho,\eta) +\left( g_1(\rho,\eta) -1 \right) 
   \cr 
   &=  C_{2b}(\rho,\rho) + \int_{\eta}^\rho \, d\eta'
   (\eta-\eta') \, e^{-\eta'(\rho-\eta')} \, C_{2b}(\rho,\eta')
   +  \left( g_1(\rho,\rho)-  1 \right) \,, 
\end{align}
from which we read off that
\begin{align} 
 g_{2b}(\rho,\rho) \equiv G_{2b}(\rho^2) &= C_{2b}(\rho,\rho) 
 +\left( C_1(\rho)- 1 \right) \,.
\end{align} 
Equating the two expressions for $g_{2b}(\rho,\eta)$, we obtain after some algebra the one-fold integral equation
\begin{align}
\label{eq:onefoldinteq_g2}
    g_{2b}(\rho,\rho) = \int_0^\rho d\rho' \rho' e^{-\rho'(\rho-\rho')} \bigg(1+g_{2b}(\rho',\rho')\bigg) + \int_0^\rho d\rho' \rho' \, {\cal K}(\rho'(\rho-\rho')) \, C_1(\rho') \,.
\end{align}
For the inhomogeneous term, the integration kernel arises from a specific integral over the Sudakov factor,
\begin{align}
     {\cal K}(x) = \int_0^\infty dx' \, \left[\frac{\theta(x-x')}{x(x-x')}\right]_+ x' e^{-x'} 
     = \frac{e^{-x}-1}{x} + e^{-x} \left( \text{Ei}(x) - \ln x - \gamma_E \right) \,,
\end{align}
where the plus-distribution subtracts the limit $x' \to x$ from the integrand, and $\text{Ei}(x)$ denotes the exponential integral, defined e.g.~via the principal value
\begin{equation}
    \text{Ei}(x) = - \text{PV} \, \int_{-x}^\infty \! dt \, \frac{e^{-t}}{t} \,.
\end{equation}
Solving the integral equation iteratively yields (for $\rho^2 = z$)
\begin{align}
\label{eq:G2smallz}
  1 + G_{2b}(z) & = \sum_{n=0}^\infty b_n \, (-z)^n
    = 1 - 2 \left(\frac{z^2}{4!} + \frac{z^3}{6!} -\frac{7z^4}{8!} + \frac{23 z^5}{10!} \right) + {\cal O}(z^6) \,.
\end{align}
Now the expansion coefficients can be shown to obey the recursive relation
\begin{align}
\label{eq:G2coefficientsrecursion}
    b_n &=  -\sum_{k=0}^{n-1} \,\frac{(n+k)!}{(2n)!} \bigg\{
    b_k - \bigg( H_{n-k-1}+\frac{1}{n-k}\bigg) a_k\bigg\}\,, \qquad (n\geq 1) \,,
\end{align} 
with $b_0 = 1$ and $b_1 = 0$, and where $H_{n}$ is the $n$-th harmonic number. 

The series expansion for the physical form factor $F(\gamma)$ in \eqref{eq:Ffinaldecomp} can then be obtained from the expansions~\eqref{eq:Gmsmallz} and~\eqref{eq:G2smallz}, by employing the relations~\eqref{eq:G1GM} as well as
\begin{equation}
    G_2(z) = G_{2a}(z) + \frac{1}{4 N_c C_F} G_{2b}(z) \,.
\end{equation}
This correctly reproduces the fixed-order expressions for the leading double logarithms derived in the previous section up to NNLO.
Furthermore, we obtain a prediction for the respective three-loop coefficient,
\begin{equation}
    F^{(3)}(\gamma) \simeq \xi_0 L^6 \left(C_F^3 \, \frac{29+44\bar{u}_0}{90\bar{u}_0^3}
    - C_F^2 C_A \, \frac{1-28\bar{u}_0}{180\bar{u}_0^3}\right) \,,
\end{equation}
which we cross-checked with an independent calculation based on pole cancellation arguments in a method-of-regions analysis~\cite{BBFHS}, as well as all higher terms in the perturbative series.

\subsection{Asymptotic behaviour in the large-recoil limit}
\label{sec:asymptotics}

It is an interesting question whether or not heavy-to-light form factors receive a Sudakov-like suppression at large hadronic recoil.
Although the actual value of the $b$-quark mass is not extremely large, the dependence of the form factor on the large energy $E_\eta \sim m_B$ in the asymptotic limit is of conceptual importance for the convergence of the heavy-quark expansion. 
In this section, we derive the asymptotic form of the soft-overlap form factor $F(\gamma)$ analytically by means of a method-of-regions analysis of the one-fold integral equations~\eqref{eq:onefoldinteqs} and~\eqref{eq:onefoldinteq_g2} for the on-shell form factors $G_1(z)$ and $G_{2b}(z)$. 
We note that in the formal limit $z \propto \alpha_s L^2 \to \infty$ it is sufficient to evaluate the coupling constant at some fixed value, because its running is a single-logarithmic effect. 

To explain our strategy, we first discuss the on-shell form factor $G_1(z)$, which obeys the simpler integral equation~\eqref{eq:onefoldinteqs_integrated}.
The key observation is that in the limit $\rho \to \infty$, the Sudakov factor in the integrand exponentially suppresses the integral in almost the entire integration domain.
The integral is thus dominated by regions in which the Sudakov exponent is not parametrically large.
An expansion in inverse powers of $z = \rho^2$ 
can then be constructed by means of a method-of-regions analysis. 
Interestingly, the mathematical structure as well as the resulting asymptotic form shares some similarities with the integrals that appear in the resummation of super-leading logarithms and the so-called Glauber series in non-global jet cross sections at hadron colliders~\cite{Becher:2023mtx,Boer:2024hzh,Boer:2024xzy} (see, for example, Sec.~4 in~\cite{Boer:2024hzh}).
Counting powers of $1/\rho \ll 1$, we decompose the integral into two relevant regions in which the exponent $\rho'(\rho-\rho') \sim {\cal O}(1)$:
\begin{align}
    &\text{region (A)} \quad \rho' \sim \rho^{-1} \,, & &\text{such that} \quad 
    \rho' (\rho-\rho') \approx \rho' \rho  \sim {\cal O}(1) 
    \nonumber \,, \\
    &\text{region (B)} \quad \rho-\rho' \sim \rho^{-1} \,, & &\text{such that}
    \quad
    \rho'(\rho-\rho') \approx \rho (\rho-\rho') \sim {\cal O}(1) \nonumber \,.
\end{align}
In region (A), the function $C_1(\rho')$ in the integrand can be expanded for small arguments, which results in a straightforward expansion in inverse powers of $\rho^2 = z$,
\begin{equation}
\frac{1}{\rho} \, \int_0^\rho d\rho'\,  e^{- \rho'(\rho-\rho')} \, C_1(\rho') \stackrel{(A)}{=} \frac{C_1(0)}{\rho} \ \int_0^\infty d\rho' \, e^{-\rho \rho'} +
     {\cal O}(z^{-2}) \approx \frac{G_1(0)}{z}
    \,,
\end{equation}
with $G_1(0) = 1$ (as well as all derivatives at $z=0$)
known from the perturbative expansion around small arguments in (\ref{eq:Gmsmallz}).
The series in region (B) results from expanding around $\rho'=\rho$ and  yields
\begin{align}
    & \frac{1}{\rho} \, \int_0^\rho \! d\rho' \,
    e^{-\rho'(\rho-\rho')} \, C_1(\rho') 
    \stackrel{(B)}{=} \,  \frac{G_1(z)-2 G_1'(z)}{z} + \frac{2G_1(z)}{z^2}
    + \mathcal{O}(G_1(z)/z^3) \,,
\end{align}
where the explicit power-counting of $G_1(z)$ and its derivatives has to be determined from the solution below in a self-consistent way.
Reinserting the sum of the two regions into the integral equation in~\eqref{eq:onefoldinteqs} gives a systematically improvable ordinary differential equation for the on-shell form factor $G_1(z)$,
\begin{align}
    0 &= - 1 + \frac{1}{z} + 
    \left(1  - 2 \, \frac{d}{dz}   \right) \frac{G_1(z)}{z} + \ldots \,,
\end{align}
which can be solved order-by-order in $1/z$, 
leading to
\begin{equation}
\label{eq:Gmasy}
    G_1(z) = z - 1 + {\cal O}(1/z)  \,.
\end{equation}

\begin{figure}[t]
    \centering
   \begin{minipage}{0.495\textwidth}
       \includegraphics[scale=0.42]{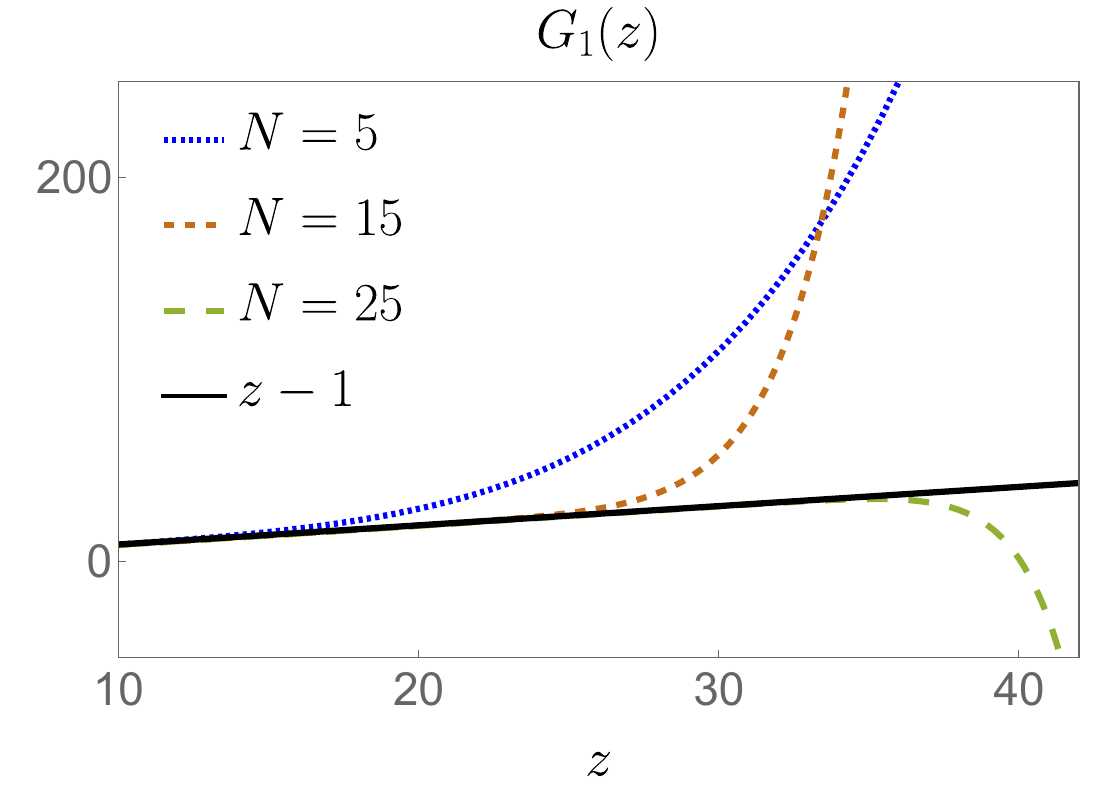}
   \end{minipage}
   \begin{minipage}{0.495\textwidth}
    \vspace{0.2mm} \includegraphics[scale=0.435]{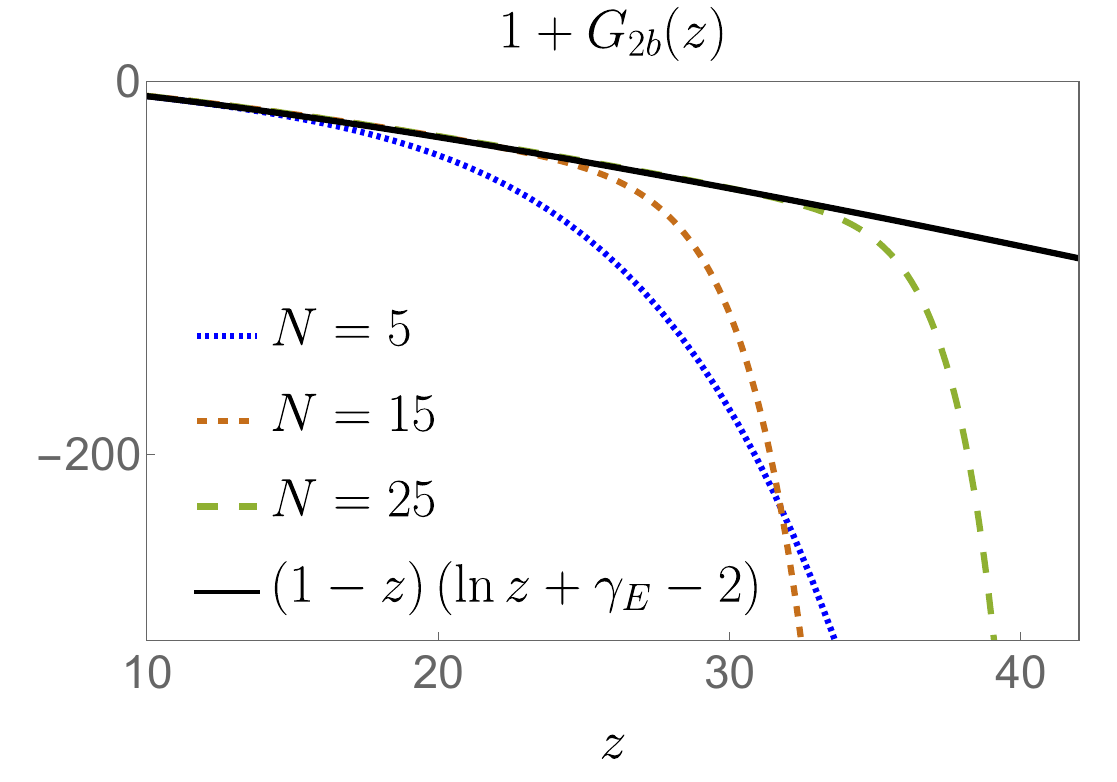} 
   \end{minipage}
   \caption{
   Partial sums of the two form factors $G_1(z;N) \equiv \sum_{n=0}^N a_n (-z)^n$ (left panel) and $1+G_{2b}(z;N) \equiv \sum_{n=0}^N b_n (-z)^n$ (right panel) as a function of $z = \frac{\alpha_s C_F}{2\pi} L^2$, and their convergence towards the asymptotic limit for large values of $z$, given in~\eqref{eq:Gmasy} and~\eqref{eq:G2asy} (black solid curves). The recursion relations that determine the coefficients $a_n$ and $b_n$ to all orders can be found in~\eqref{eq:Gmcoefficientsrecursion} and~\eqref{eq:G2coefficientsrecursion}, respectively.}
   \label{fig:largezasymptotics}
\end{figure}

The fact that this result is so simple is actually quite interesting, and results from a delicate interplay of soft-quark and soft-gluon corrections. To understand this, recall that the expression for the resummed pure soft-quark logarithms in~\eqref{eq:puresoftquarks} is a linear combination of modified Bessel functions, which grow exponentially for asymptotically large arguments,
\begin{equation}
    I_n(2\sqrt{z}) \approx \frac{1}{2\sqrt{\pi}} \frac{e^{2\sqrt{z}}}{z^{1/4}}  \,, \qquad \text{for} \quad z\to \infty \,,
\end{equation} 
whereas the Sudakov factor from soft-gluon logarithms causes an exponential suppression.
In~\eqref{eq:pdes} this is reflected by the second and third term in the differential operator.
They come with the same prefactor but opposite sign, 
which leads to a cancellation of the two effects and results in the simple linear growth for large $z$.
This is also worth mentioning because the coefficient multiplying the soft-gluon logarithms can be identified with the (leading-order) cusp anomalous dimension. Our results therefore suggest that -- at least at the leading double-logarithmic level -- also the soft-quark logarithms come with the same coefficient. We mention a possible explanation of this observation later in Sec.~\ref{sec:discussion}.

\begin{figure}[t]
    \centering
   \includegraphics[width=0.6\textwidth]{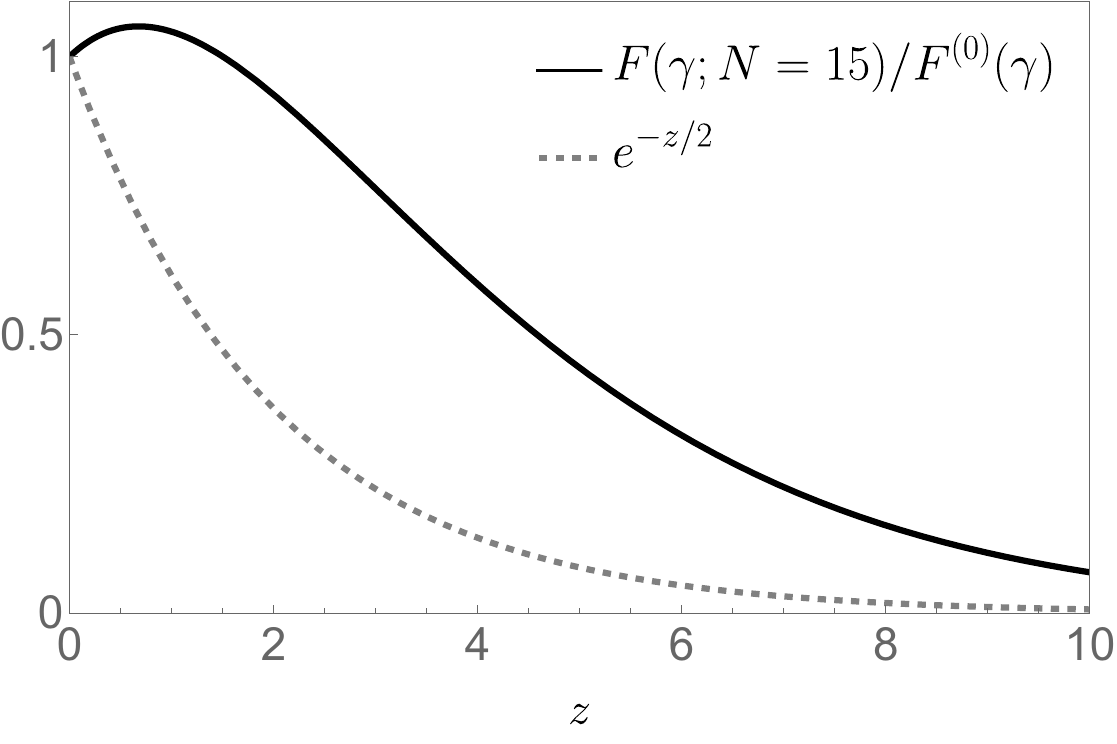}
    \caption{Comparison of the large-$z$ fall-off between the standard multiplicative Sudakov factor in the double-logarithmic approximation (dashed curve) and the resummed expression for the physical form factor $F(\gamma)$ (solid curve).
    Both curves are normalized to the Born expression $F^{(0)}(\gamma)$, and for the solid curve we evaluate the partial sum $F(\gamma;N) = \sum_{n=0}^N \left(\frac{\alpha_s}{4\pi}\right)^n F^{(n)}(\gamma)$ for $N=15$ with equal quark masses, $\bar{u}_0 = 1/2$.
    As the plots from Fig.~\ref{fig:largezasymptotics} show, the black curve is well approximated by this polynomial for $z \leq 10$.
    We recall that the variable $z = \frac{\alpha_s C_F}{2\pi} L^2$ is related to the large boost $\gamma$ through $L = \ln(2\gamma)$.}
    \label{fig:comparison_with_Sudakov}
\end{figure}

The discussion of the asymptotics of the form factor $G_{2b}(z)$ is slightly more subtle due to the inhomogeneous mixing term in~\eqref{eq:onefoldinteq_g2}.
Here the integration kernel ${\cal K}(x) \simeq \frac{1}{x^2} + \frac{2}{x^3} + \dots$ has only a power-like fall-off for large arguments.
As a consequence, the integral receives contributions from two additional regions:
\begin{align}
    &\text{region (C)} \quad \rho' \sim \rho-\rho' \sim \rho \,, & &\text{such that} \quad 
    \rho' (\rho-\rho') \gg 1
    \nonumber \,, \\
    &\text{region (D)} \quad \rho' \sim {\cal O}(1) \,, & &\text{such that}
    \quad
    \rho'(\rho-\rho') \approx \rho' \rho \gg 1 \nonumber \,.
\end{align}
Furthermore, it turns out that the integrals in all regions are logarithmically divergent, and only their sum is finite.
As a result, the asymptotic expansion of the form factor $G_{2b}(z)$ receives a logarithmic dependence on $z$ from the inhomogeneous term,
\begin{equation}
    \int_0^\rho d\rho' \rho' {\cal K}\big( \rho' (\rho-\rho')\big) \, C_1(\rho') 
     \simeq - \ln z - \gamma_E -1 + \frac{1}{z} \left(-\ln z - \gamma_E + c \right) + \mathcal{O}(z^{-2}) \,.
\end{equation}
As region~(D) contains an integral over the entire function $C_1(\rho')$, the above result contains a constant $c$ that cannot
be determined analytically from the known expansion of $C_1(\rho')$ for small and large arguments. However, when numerically integrating the series expansion we find $c\simeq 1$ at the level of $\mathcal{O}(10^{-4})$.
Eventually, after performing the same steps as for the form factor $G_1(z)$, we obtain
\begin{align}
\label{eq:G2asy}
    1+G_{2b}(z) &= (1-z) \left( \ln z + \gamma_E-2 \right)  + {\cal O}(1/z) \,.
\end{align}
The formally dominant term in the limit $E_\eta \sim m_b \to \infty$ is thus $\sim z \ln z$, but we remind the reader that the contribution of $G_{2b}(z)$ to the physical form factor is suppressed by $1/N_c^2$ in the large $N_c$ limit.

In Fig.~\ref{fig:largezasymptotics} we illustrate the convergence of the perturbative series for the form factors $G_1(z)$ and $1+G_{2b}(z)$ towards their asymptotic expansion for large values of $z$, which shows that already for moderate values of the truncation order $N$ the fixed-order expansion approximates its asymptotic limit well. We also compare the large-$z$ behaviour of the physical form factor $F(\gamma)$ with the standard Sudakov suppression factor in the double-logarithmic approximation $e^{-z/2}$ in Fig.~\ref{fig:comparison_with_Sudakov}. 
Recall that the latter function only describes the resummation of soft-gluon logarithms from attachments at the heavy-to-light vertex, ignoring their interplay with soft-quark configurations. As can be observed in the figure, the soft-quark corrections somewhat weaken the fall-off  for large values of $z = \frac{\alpha_s C_F}{2\pi} L^2$ by a linear term.

\section{Discussion}
\label{sec:discussion}

One of the central results of our paper are the coupled integral equations in~\eqref{eq:inteqslogvar}, which reflect the iteration of rapidity-ordered soft (on-shell) quarks in the presence of resummed soft-gluon corrections. As already mentioned, the dynamical origin of the iterative structure is the same as for energetic muon-electron backward scattering discussed in \cite{Bell:2022ott}. In that case, it has been shown in the context of a bare factorization theorem in SCET that the nested longitudinal momentum integrals lead to an iterative structure of endpoint singularities. More precisely, the refactorization conditions describing the endpoint behaviour of (bare) soft and collinear functions are implicit integral equations that again contain endpoint-divergent convolutions. In contrast to other power-suppressed processes, for which the problem of endpoint-divergent convolutions could be solved by an additive rearrangement of the terms in the factorzation formula (see e.g.~\cite{Beneke:2020ibj,Beneke:2022obx, Liu:2019oav,Hurth:2023paz}), this nested structure of endpoint singularities currently prevents a consistent and complete factorization of the soft and collinear dynamics in an EFT framework. The same conclusion applies for the soft-overlap contribution to $B_c\to \eta_c$ form factors analyzed in this work (see also~\cite{Boer:2018mgl}). 

We should also mention that the situation becomes even more involved in the case of $B$-decays to light mesons, where the non-perturbative bound-state effects have to be included in terms of hadronic matrix elements of properly defined operators in the EFT. Although the massless-quark limit is non-trivial, we believe that our results provide valuable constraints on the properties of soft and collinear functions that one would have to define in the context of QCD factorization and SCET. As an example, we note an interesting correspondence between the partial differential equations~\eqref{eq:pdes} following from \eqref{eq:inteqslogvar} and the renormalization-group equations for endpoint-divergent inverse moments of light-cone distribution amplitudes, as they would appear in the QCD factorization approach. Considering, for instance, the well-known Efremov-Radyushkin-Brodsky-Lepage evolution kernel~\cite{Lepage:1979zb,Lepage:1980fj,Efremov:1979qk} for the leading-twist light-cone distribution amplitude $\phi(u;\mu)$ of a light pseudoscalar meson in the limit $u\to 0$~\cite{Beneke:2021pkl}, and keeping only logarithmically enhanced terms in $\ln u$, the one-loop evolution equation becomes 
\begin{equation}
\label{eq:asymptoticRGE}
    \frac{d}{d\ln\mu^2} \, \phi(u;\mu) \simeq \frac{\alpha_s C_F}{2\pi} \bigg[ \phi(u;\mu) \, \ln u + u \int_u^1 \frac{dv}{v^2} \, \phi(v;\mu) \bigg] \,.
\end{equation}
For $u \to 0$ the momentum of the spectator antiquark becomes soft, whereas the active quark carries almost the entire energy of the collinear meson. As discussed in~\cite{Beneke:2021pkl}, the one-loop evolution kernel for a light pseudoscalar meson then essentially reduces (up to a constant) to the Lange-Neubert kernel for the leading-twist $B$-meson light-cone distribution amplitude~\cite{Lange:2003ff}. As the latter is of Sudakov type, it features terms proportional to the cusp anomalous dimension, and it turns out that both terms in~\eqref{eq:asymptoticRGE} are associated with its leading-order coefficient. 

As we shall see now, this may explain the observation made before, that the double logarithms from soft-gluon \emph{and} soft-quark configurations come with the same prefactor. To this end, defining a regularized inverse moment of the leading-twist distribution amplitude with an explicit lower cutoff,
\begin{equation} \langle u^{-2} \rangle(\kappa,\mu) \equiv \int_\kappa^1 \frac{du}{u^2} \; \phi(u,\mu) \,, 
\end{equation}
one finds that this object fulfills a differential equation in the renormalization scale $\mu$ and the endpoint cutoff $\kappa$, which is given by
\begin{equation}
 \bigg[ \frac{d^2}{d\ln\kappa \ d\ln\mu^2} - \frac{\alpha_s C_F}{2\pi} \Big(\ln \kappa \, \frac{d}{d\ln \kappa} - 1 \Big)\bigg] \, \langle u^{-2} \rangle(\kappa;\mu) = 0 \,.
 \label{eq:RGE:cutoffmoment}
\end{equation}
Note that this has precisely the form of the differential equation for $g_1(\rho,\eta)$ in~\eqref{eq:pdes} upon identifying $\rho \propto \ln\mu^2$ and $\eta \propto -\ln \kappa$ without using any specific information about the non-relativistic bound states. In other words, the solution of \eqref{eq:RGE:cutoffmoment} encodes the simultaneous resummation of large logarithms $\ln\mu^2/\mu_0^2$ and $\ln u$, which determine the dominant contribution to the cutoff moment in the limit $\kappa \sim 1/\gamma \ll 1$. However, we emphasize that at this point it is still an open question how to systematically derive a factorization theorem that is free of endpoint singularities, starting from a bare factorization formula. Further details on the origin of such correspondences between the diagrammatic analysis and the formulation in terms of (bare) factorization theorems in SCET will be discussed in a future publication~\cite{BBFHS}.

\section{Conclusion}
\label{sec:conclusion}

In this article, we addressed the so-called ``soft-overlap'' contribution to heavy-to-light transition form factors at large recoil within a non-relativistic framework. Specifically, we focused on the decay $B_c \to \eta_c$ in the limit $m_b \gg m_c \gg \Lambda_{\rm QCD}$, in which the hadronic matrix elements become accessible in QCD perturbation theory. We performed a detailed study of the leading logarithmically enhanced one-loop and two-loop corrections to the ``soft'' form factor $F(\gamma)$, where $\gamma$ denotes the large kinematic boost of the final-state meson $\eta_c$ relative to the $B_c$ rest frame. A systematic analysis of the relevant momentum configurations in various Feynman diagrams reveals two distinct sources of large double logarithms with an intricate interplay:
\begin{itemize}
\item 
Soft-quark logarithms emerge from rapidity-ordered (on-shell) quark propagators, and lead to nested integrals in the longitudinal light-cone momenta. Their structure resembles the situation of energetic muon-electron backward scattering studied by some of us in~\cite{Bell:2022ott}, but here we find the additional complication of mixing between various Dirac structures, as well as contributions from non-Abelian interactions.
\item    
The exponentiation of soft-gluon logarithms results in well-known Sudakov factors, which depend on the soft-quark momenta, and therefore modify the structure of the nested soft-quark integrals in a non-trivial way.
\end{itemize}
As we have shown in this work, the simultaneous resummation of the leading double logarithms $\alpha_s^n \ln^{2n} (2\gamma)$ to all orders from both sources is governed by a set of two coupled implicit integral equations, given in \eqref{eq:finalintegralequations}. To the best of our knowledge, the structure of these equations is of a novel type, and has not been discussed in the context of exclusive decay amplitudes in the literature so far. While special cases lead to the well-known solutions for the bottom-induced $h\to\gamma\gamma$ decay and muon-electron backward scattering, a closed-form solution of these equations remains at present unknown. Instead we showed that the perturbative expansion can be determined iteratively in a straightforward manner to practically any order. Notably, this reproduces the independently computed fixed-order results up to three-loop order, and further predicts the leading double logarithms of the form factor to all orders. We furthermore derived the asymptotic behavior of the soft-overlap  form factor $F(\gamma)$ in the formal limit $ z \propto \alpha_s \ln^2 (2\gamma) \to \infty$, which shows that its overall Sudakov suppression is somewhat weakened by soft-quark effects through an additional linear term in $z$.

In conclusion, our analysis provides new insights into the structure of logarithmically enhanced corrections to heavy-to-light form factors, and imposes valuable constraints on the long-standing goal of understanding soft-collinear factorization for power-suppressed exclusive processes. A dedicated discussion of our results in the context of QCD factorization, as well as a detailed comparison with the fixed-order three-loop computation  will be presented in a forthcoming publication~\cite{BBFHS}.

\subsubsection*{Acknowledgements}

The research of GB, TF, DH and VS was supported by the Deutsche Forschungsgemeinschaft (DFG, German Research Foundation) under grant 396021762 - TRR 257. The research of PB was supported by the  Cluster  of  Excellence  ``Precision  Physics,  Fundamental Interactions, and Structure of Matter" (PRISMA$^+$ EXC 2118/1) funded by the German Research Foundation (DFG) within the German Excellence Strategy (Project ID 390831469), and was funded by the European Union's Horizon 2020 Research and Innovation Program under the Marie Sk\l{}odowska-Curie grant agreement No.101146976 and the European Research Council (ERC) under the European Union’s Horizon 2022 Research and Innovation Program (ERC Advanced Grant agreement No.101097780, EFT4jets). GB and PB thank the Erwin-Schrödinger International Institute for Mathematics and Physics at the University of Vienna and the Munich Institute for Astro-, Particle and BioPhysics (MIAPbP) for support. All Feynman diagrams have been produced with \texttt{FeynGame} \cite{Harlander:2020cyh, Harlander:2024qbn}.

\newpage
\bibliography{refs}
\end{document}